\documentclass[aps,pra,twocolumn,showpacs,notitlepage,superscriptaddress,letterpaper]{revtex4-1}
\usepackage{graphicx}
\usepackage{amsmath}
\usepackage{esint}
\usepackage{verbatim}
\usepackage{color}
\usepackage{SIunits}
\usepackage{hyperref}

\newcommand{\nbath}{n_{{b}}}

\newcommand{\Tbath}{T_{b}}
\newcommand{\Tf}{T_{f}}

\newcommand{\Tm}{T_{m}}

\newcommand{\kappaEven}{\kappa_+}
\newcommand{\kappaeEven}{\kappa_{e,+}}
\newcommand{\kappaiEven}{\kappa_{i,+}}
\newcommand{\kappaOdd}{\kappa_-}
\newcommand{\kappaeOdd}{\kappa_{e,-}}
\newcommand{\kappaiOdd}{\kappa_{i,-}}
\newcommand{\omegac}{\omega_{r,0}}
\newcommand{\omegacEven}{\omega_{r,+}} 
\newcommand{\omegacOdd}{\omega_{r,-}}

\newcommand{\GOM}{G}

\newcommand{\Cm}{C_{m}}
\newcommand{\Cs}{C_{s}}

\newcommand{\Ctot}{C_{\text{tot}}}

\newcommand{\Ceff}{C_\text{eff}}

\newcommand{\ahat}{\hat{a}}
\newcommand{\adag}{\hat{a}^{\dagger}}
\newcommand{\bhat}{\hat{b}}
\newcommand{\bdag}{\hat{b}^{\dagger}}
\newcommand{\ain}{\hat{a}_\text{in}}

\newcommand{\bb}{\hat{b}_b}
\newcommand{\aout}{\hat{a}_\text{out}}
\newcommand{\abOdd}{\hat{a}_{b,r-}}
\newcommand{\abEven}{\hat{a}_{b,r+}}

\newcommand{\chirEven}{\chi_{r+}}
\newcommand{\chirOdd}{\chi_{r-}}

\newcommand{\Wn}{{W}_{n}} 
\newcommand{\an}{{a}_{n}}
\newcommand{\Wx}{{W}_{x}}
\newcommand{\Wy}{{W}_{y}} 
\newcommand{\tAl}{t_\text{Al}} 
\newcommand{\tSi}{t_\text{Si}} 
\newcommand{\xzpf}{x_{\text{zpf}}} 
\newcommand{\meff}{m_{\text{eff}}}
\newcommand{\am}{{a}_{m}}
\newcommand{\Wm}{{W}_{m}}
\newcommand{\hm}{{h}_{m}}

\newcommand{\gammai}{\gamma_{i}}
\newcommand{\omegam}{\omega_{m}}

\newcommand{\gammam}{\gamma_{m}}
\newcommand{\gammaem}{\gamma_\text{em}}

\newcommand{\gammap}{\gamma_p}
\newcommand{\chim}{\chi_{m}}

\newcommand{\Hem}{\hat{H}_\text{em}}

\newcommand{\gzeroE}{g_{\text{0}}}
\newcommand{\gzeroEeff}{g_{\text{0},\pm}}

\newcommand{\omegap}{\omega_{p}}
\newcommand{\omegad}{\omega_{d}}
\newcommand{\ndriveOdd}{n_{d,-}}
\newcommand{\ndriveEven}{n_{d,+}}
\newcommand{\Pdrive}{P_{d}}
\newcommand{\DeltardOdd}{\Delta_{r-,d}}
\newcommand{\DeltardEven}{\Delta_{r+,d}}
\newcommand{\deltarpEven}{\delta_{r+,p}}

\newcommand{\nmbath}{n_{\text{b,m}}}

\newcommand{\nrEven}{n_{r,+}}

\newcommand{\nm}{n_{m}}

\newcommand{\nbEven}{n_{b,r+}}
\newcommand{\nmpump}{n_{m,p}}

\begin{document}

\title{Quantum electromechanics of a hypersonic crystal}

\author{Mahmoud Kalaee}
\thanks{These authors contributed equally to this work.}

\author{Mohammad Mirhosseini}
\thanks{These authors contributed equally to this work.}

\author{Paul B. Dieterle}
\affiliation{Kavli Nanoscience Institute and Thomas J. Watson, Sr., Laboratory of Applied Physics, California Institute of Technology, Pasadena, CA 91125, USA}
\affiliation{Institute for Quantum Information and Matter, California Institute of Technology, Pasadena, CA 91125, USA}

\author{Matilda Peruzzo}
\affiliation{Institute of Science and Technology Austria, 3400, Klosterneuburg, Austria}

\author{Johannes M. Fink}
\affiliation{Institute of Science and Technology Austria, 3400, Klosterneuburg, Austria}
\affiliation{Institute for Quantum Information and Matter, California Institute of Technology, Pasadena, CA 91125, USA}

\author{Oskar Painter}
\email{opainter@caltech.edu}
\affiliation{Kavli Nanoscience Institute and Thomas J. Watson, Sr., Laboratory of Applied Physics, California Institute of Technology, Pasadena, CA 91125, USA}
\affiliation{Institute for Quantum Information and Matter, California Institute of Technology, Pasadena, CA 91125, USA}  

\date{\today}
\begin{abstract}
  Radiation pressure within engineered structures has recently been used to couple the motion of nanomechanical objects with high sensitivity to optical and microwave electromagnetic fields. Here, we demonstrate a form of electromechanical crystal for coupling microwave photons and hypersonic phonons by embedding the vacuum-gap capacitor of a superconducting resonator within a phononic crystal acoustic cavity.  Utilizing a two-photon resonance condition for efficient microwave pumping and a phononic bandgap shield to eliminate acoustic radiation, we demonstrate large cooperative coupling ($C \approx 30$) between a pair of electrical resonances at $\omegac \approx 10$~GHz and an acoustic resonance at $\omegam/2\pi = 0.425$~GHz.  Electrical read-out of the phonon occupancy shows that the hypersonic acoustic mode has an intrinsic energy decay time of $2.3$~ms and thermalizes close to its quantum ground-state of motion (occupancy $\nm = 1.5$) at a fridge temperature of $\Tf = 10$~mK.  Such an electromechanical transducer is envisioned as part of a hybrid quantum circuit architecture, capable of interfacing to both superconducting qubits and optical photons.

\end{abstract}
\maketitle

Superconducting electromechanical circuits have recently been used to demonstrate exquisitely sensitive measurement and control of mesoscopic mechanical objects in the quantum regime~\cite{Teufel2011b,Teufel2011a,Pirkkalainen2013}.  In these devices, mechanical coupling is typically via a dispersive interaction with a high $Q$-factor electromagnetic resonator, whereby mechanical motion phase modulates the internal electromagnetic field resulting in the generation of motional sidebands.  This interaction is parametric, allowing for the coupling of low frequency mechanical oscillators to much higher frequency electromagnetic resonators, and can be enhanced through application of a strong electrical driving tone.  The strength of the coupling at the quantum level is defined by a vacuum rate, $\gzeroE$, which for capacitively coupled circuits is related to the scale of the mechanical quantum zero-point motion in comparison to the dimension of the capacitor.  Superconducting microwave resonators employing nanoscale vacuum-gap capacitors can reach vacuum coupling levels as large as a few hundred Hz to MHz-frequency mechanical oscillators~\cite{Teufel2009}, and have been utilized for a variety of applications ranging from conversion between microwave and optical photons~\cite{Safavi2011,Andrews2014} to the generation and detection of squeezed states of mechanical motion~\cite{Wollman2015,Pirkkalainen2015,Lecocq2015a}.

Similar work in the optical domain has sought to increase the radiation pressure within optical resonators by scaling the optical mode volume down to the nanoscale~\cite{Aspelmeyer2014}.  An example of this is the optomechanical crystal (OMC)~\cite{Eichenfield2009}, in which large coupling between near-infrared photons and hypersonic ($\gtrsim$ GHz)~\cite{Maldovan2013} phonons has been realized.  Owing to the factor of $\sim 10^5$ between the speed of light and sound in solid-state materials, optical photons and hypersonic phonons are matched in wavelength, enabling the construction of integrated photonic and phononic circuits which can be used to route signals around on a chip or to inter-convert optical and acoustic waves~\cite{Balram2016,Fang2016,Fang2017,Patel2018}.  For quantum applications, hypersonic acoustic devices also have several advantages.  GHz-level frequencies facilitate operation in the sideband-resolved limit of optomechanics~\cite{Aspelmeyer2014}, a crucial parameter regime for realizing noise-free quantum signal conversion~\cite{Safavi2011,Hill2012}.  Additionally, the integration of superconducting quantum circuits~\cite{Schoelkopf2008} and microwave acoustic devices is actively being explored~\cite{LaHaye2009,Oconnell2010,Pirkkalainen2013,Gustafsson2014,Chu2017,Manenti2017,Satzinger2018,Moores2018,Arrangoiz_Arriola2018}, where the compact wavelength and lack of fringing fields in vacuum of acoustic phonons can enable superior miniaturization and scaling.   

Towards this effort, the use of piezoelectric materials~\cite{Eom2012b,Piazza2012} can enable MHz-rate electromechanical coupling suitable for quantum information processing~\cite{Balram2016,Han2016,Arrangoiz_Arriola2016,Chu2017,Manenti2017,Satzinger2018,Moores2018,Arrangoiz_Arriola2018}. The piezoelectric coupling, however, cannot be turned off nor is it perfectly mode selective, and poly-crystalline piezoelectric materials can harbor lossy defects~\cite{Phillips1972}.  Both these effects can lead to parasitic electrical or acoustic decoherence.  Parametric radiation pressure coupling can be dynamically controlled and is relatively materials agnostic, but it is challenging to reach the requisite level of coupling due to the large mismatch in electromagnetic and acoustic wavelengths.  Using an aluminum (Al) on silicon-on-insulator (SOI) process which has been effective in forming low-loss superconducting quantum circuits~\cite{Keller2017}, here we demonstrate an electromechanical resonator that utilizes hypersound frequency phononic crystals to engineer the localization and parametric coupling of mechanical motion at $\omegam/2\pi=0.425$ GHz to an integrated superconducting microwave high impedance circuit.  This electromechanical crystal (EMC) structure, akin to the optical OMCs, achieves simultaneously the large photon-phonon coupling ($\gzeroE/2\pi=17.3$~Hz) and minimal acoustic damping ($\gammai/2\pi = 68$~Hz) required of quantum electromechanics applications.

\vspace{2mm}\noindent\textbf{Electromechanical crystal design and fabrication}

\begin{figure}[t!]
\begin{center}
\includegraphics[width=\columnwidth]{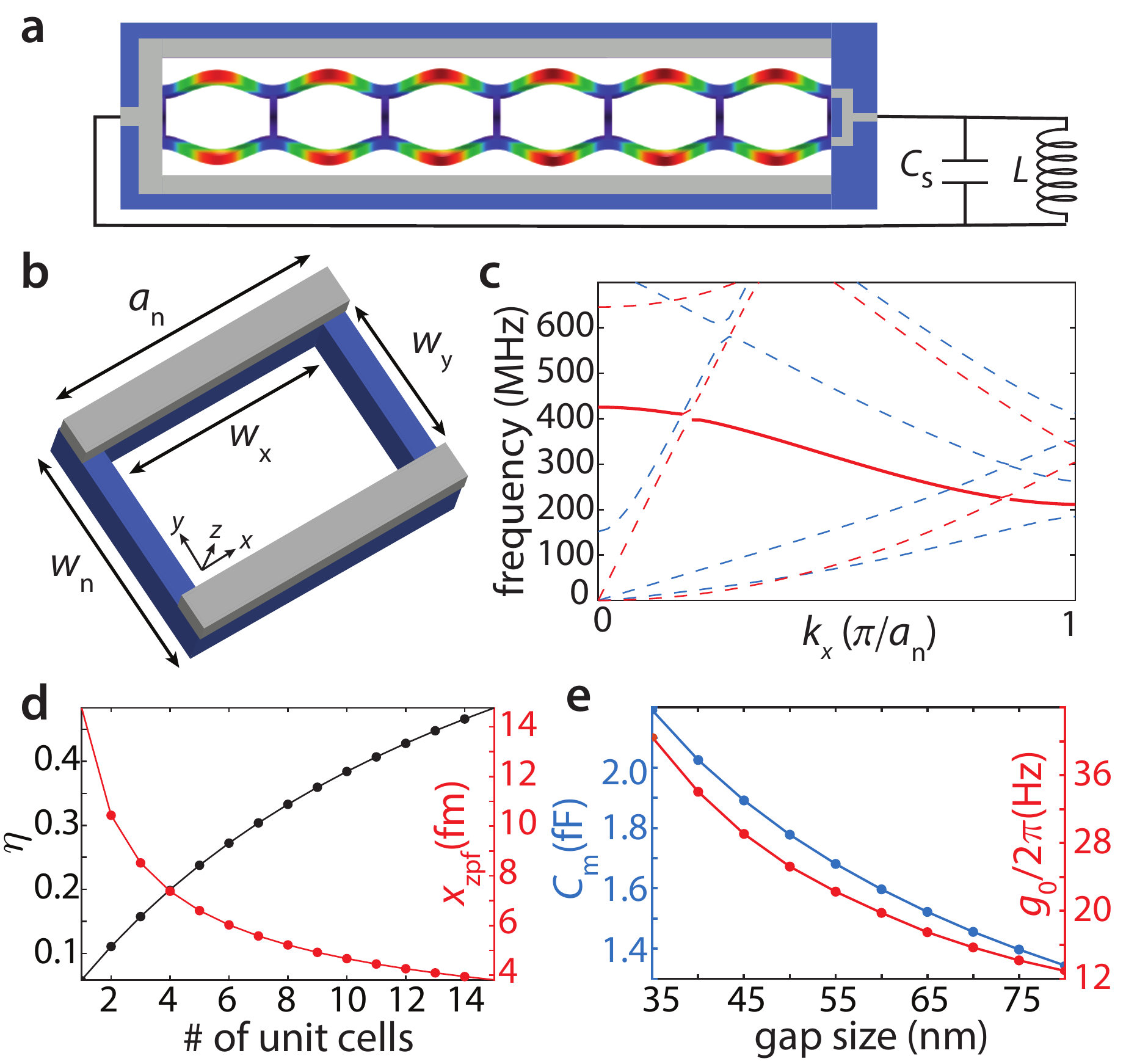}
\caption{\textbf{Nanobeam phononic crystal design.} \textbf{a}, Schematic of the central nanobeam region showing the breathing mode.  Mechanical motion is indicated by an exaggerated displacement of the beam structure, with red (blue) color indicating regions of large (small) amplitude of the motion. The Al capacitor electrodes (grey) are connected in parallel to a coil inductor of inductance $L$ and parasitic capacitance $C_s$. \textbf{b}, Unit cell of the nanobeam phononic crystal lattice with Si device layer (Al electrode) shown as blue (grey).  \textbf{c}, Acoustic band structure for an infinitely periodic nanobeam phononic crystal with parameters: $\an=1.55$~$\mu$m, $\Wn=900$~nm, $\Wx=600$~nm and $\Wy=1.45$~$\mu$m. The Si device layer and Al electrode thicknesses are $\tSi=220$~nm and $\tAl=60$~nm, respectively. The red and blue curves correspond to symmetric and anti-symmetric modes with respect to the $x$-$z$ symmetry plane.  The band from which the breathing mode is formed is shown as a solid red curve. \textbf{d}, Participation ratio ($\eta$) and zero-point motion amplitude ($\xzpf$) of the breathing mode as a function of number of unit cells in the beam for a fixed parasitic capacitance $\Cs=3.1$~fF and a vacuum gap size of $45$~nm. \textbf{e}, Motional capacitance, $\Cm$, and zero-point coupling, $\gzeroE$, of the electrical circuit as a function of the gap size.  Here the coil inductance, $L$, is adjusted for each gap to keep the LC-resonance frequency fixed at $\omegac=10.77$ GHz.  See App.~\ref{app:A} and \ref{app:B} for details of mechanical and electrical numerical simulations.
}\label{fig:fig1} 
\end{center}
\end{figure}

The electromechanical crystal studied in this work is formed from superconducting Al wiring on a patterned sub-micron thick silicon (Si) membrane, and consists of three primary elements: (i) a central nanobeam phononic crystal cavity and capacitor with nanoscale vacuum gap, (ii) an acoustic shield with a phononic bandgap for all polarizations and propagation directions of acoustic waves, and (iii) a nanoscale-pitch spiral coil inductor with minimal stray capacitance and large intrinsic impedance.  Details of the planar spiral inductor are described in App.~\ref{app:A}.  Here we focus on the design of the nanobeam cavity and acoustic shield.  Figure~\ref{fig:fig1}(a) depicts the patterned nanobeam cavity geometry and Al wiring layout of the vacuum-gap capacitor.  The resulting hypersonic `breathing' acoustic cavity mode is also shown, visualized as an exaggerated deformation of the beam structure.  Referring to the nanobeam unit cell and acoustic bandstructure of Figs.~\ref{fig:fig1}(b-c), this breathing mode is formed from an acoustic band (solid bold red curve) near the $\Gamma$-point at wavevector $k_{x}=0$.  For a lattice constant of $\an=1.55$~$\mu$m numerical finite-element method (FEM) simulations place the $\Gamma$-point frequency of the breathing mode band at $\omegam/2\pi = 0.425$~GHz.  Although other acoustic bands (dashed curves) are also present, the relative isolation of the breathing $\Gamma$-point modes in reciprocal space still allows for the formation of highly localized cavity modes near the band-edge.           

Several subtle features of the nanobeam design are key to realizing large electromechanical coupling, the magnitude of which is given by~\cite{Fink2016},

\begin{align}
\gzeroE=-\eta \xzpf \frac{\omegac}{2 \Cm}\frac{\partial \Cm}{\partial u},
\end{align}

\noindent where $\omegac$ is the resonance frequency of the coupled LC circuit, $u$ is the displacement amplitude of the acoustic mode of interest, $\xzpf$ is the zero-point amplitude of this mode, and $\Cm$ is the vacuum-gap capacitance affected by the beams motion.  $\eta$ is a motional participation ratio defined by $\eta=\Cm/\Ctot$, where $\Cs$ is the stray and $\Ctot = \Cs+\Cm$ the total capacitance of the LC circuit.  Firstly, a minimum motional mass ($\meff$) is desired to increase the zero-point amplitude ($\xzpf=[\hbar/2\meff\omegam]^{1/2}$).  In the case of the patterned nanobeam this corresponds to minimizing the thickness of the Si and Al layers and minimizing the width of the beam features.  Secondly, a large motional capacitance is desired due to limits on the achievable stray capacitance.  Owing to the use of a $\Gamma$-point acoustic mode the electromechanical coupling from each unit cell is additive and increasing the number of unit cells in the acoustic cavity results in an increased motional capacitance and participation ratio.  FEM simulations of $\xzpf$ and $\eta$ versus the number of nanobeam unit cells are shown in Fig.~\ref{fig:fig1}(d) for a stray capacitance $\Cs = 3.1$~fF and a fixed vacuum gap $s=45$~nm.  Here, $\Cs$ is dominated by the stray capacitance of the planar spiral coil inductor forming the LC resonator.  Figure~\ref{fig:fig1}(e) shows the resulting simulated vacuum coupling rate versus gap size of the breathing mode for a nanobeam structure consisting of $11$ unit cells.  Beyond $11$ unit cells we find the acoustic mode becomes too sensitive to disorder, and tends to breaks up into localized resonances when fabricated.    

\begin{figure}[t!]
\begin{center}
\includegraphics[width=\columnwidth]{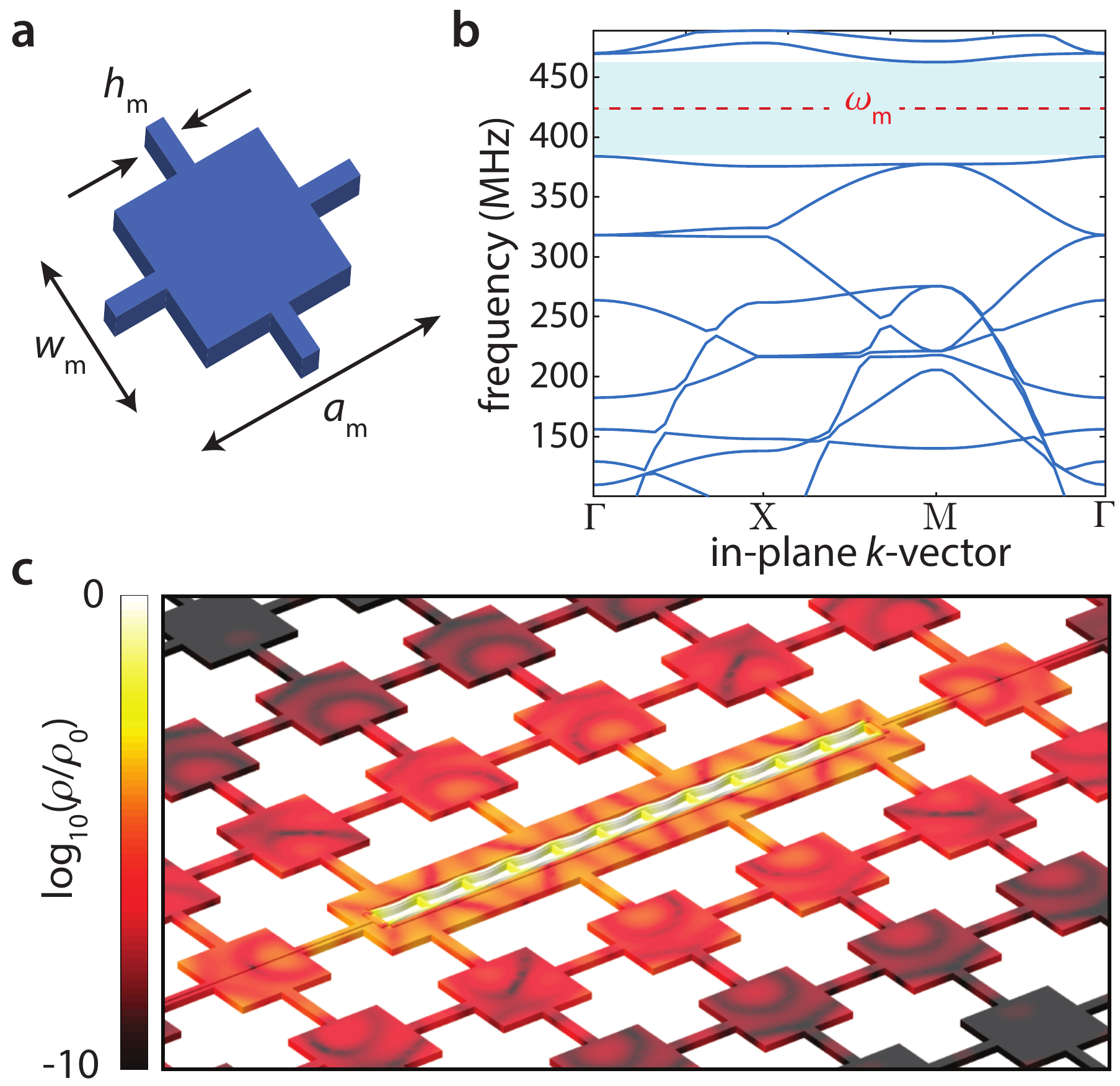}
\caption{\textbf{Phononic crystal shield.} \textbf{a}, Schematic and dimensions of a unit cell in the phononic crystal shield (mirror cell) surrounding the nanobeam central cavity. \textbf{b}, Acoustic band structure of the acoustic shield for mirror unit cell parameters: $\am=5.13~\mu$m, $\Wm=2.1~\mu$m, $\hm=360$ nm and $t_{\text{Si}}=220$~nm. The acoustic band gap is shaded in blue and the localized breathing mode frequency is marked with a dashed red line. \textbf{c}, Logarithmic scale color plot of the acoustic energy density for the nanobeam breathing of (a) embedded in the acoustic shield of (b).  Acoustic energy density, $\rho$, is normalized to its peak value, $\rho_{0}$, located in the nanobeam.  Displacement of the structure is also used to visualize the breathing mode profile.}
\label{fig:fig2}
\end{center}
\end{figure}

As mentioned above, the nanobeam phononic crystal does not have a full phononic bandgap in the vicinity of the breathing mode.  In order to provide additional acoustic isolation from the surrounding Si membrane and substrate the nanobeam cavity and vacuum-gap capacitor are embedded in the middle of a `cross-pattern' phononic bandgap crystal~\cite{Safavi-Naeini2010b}.  A unit cell of the cross shield, shown in Fig.~\ref{fig:fig2}(a), consists of a large square plate region with four narrow connecting tethers.  Through adjustment in the width of the square plate, and length and width of the connecting tethers, a substantial bandgap can be opened up between the low frequency tether modes and the localized modes of each square plate.  A FEM-simulated acoustic bandstructure of an optimized cross structure is shown in Fig.~\ref{fig:fig2}(b), where a bandgap of nearly $0.1$~GHz around the breathing mode frequency of $0.425$~GHz is obtained.  Embedding the nanobeam phononic crystal cavity in the middle of a cross phononic crystal, Fig.~\ref{fig:fig2}(c) shows a simulation of the resulting radiation pattern of the localized breathing mode.  As can be clearly seen, the energy density of the breathing mode reduces dramatically upon entering the acoustic shield, dropping by $100$~dB in only two periods.         

\begin{figure*}[ht!]
\begin{center}
\includegraphics[width=\textwidth]{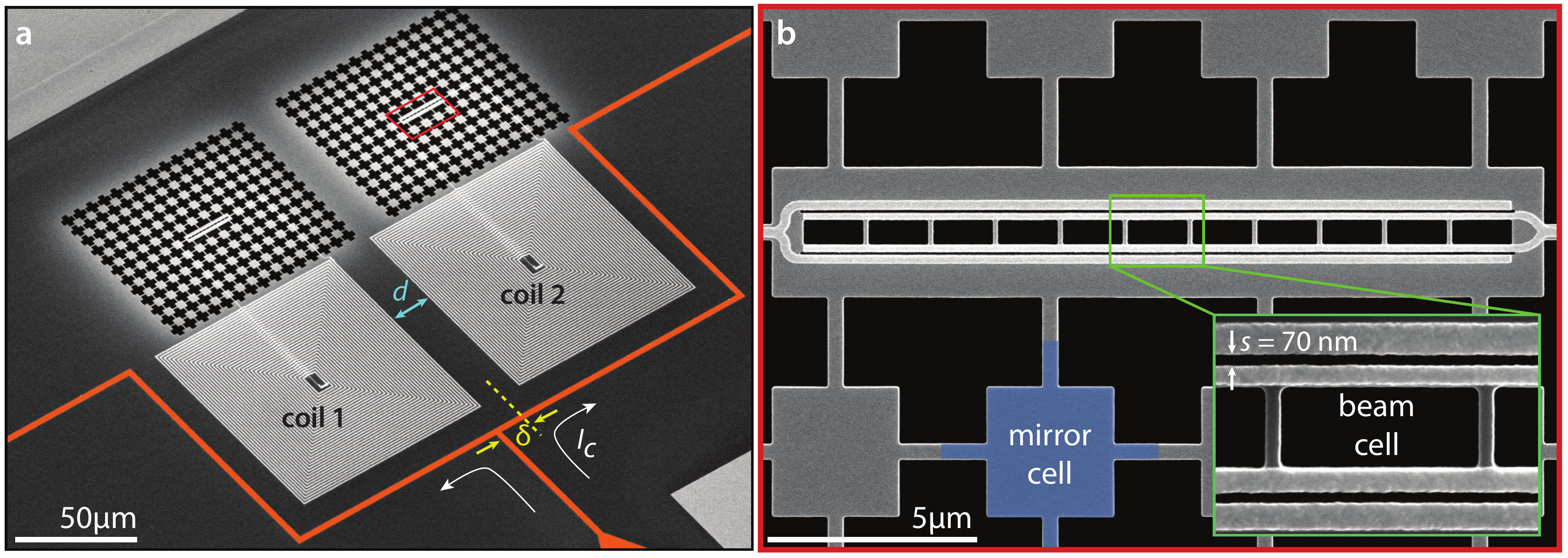}
\caption{\textbf{Fabricated structure.} \textbf{a}, Zoomed-out SEM image of the entire double-cavity device fabricated on a SOI membrane.  The two Al spiral coils, labeled coil 1 and coil 2, are shunted via the vacuum-gap capacitor of two independent nanobeam acoustic cavities which are embedded inside acoustic shields.  Inductive coupling between the two coils is set by the gap $d$.  External coupling of the two coils to a $50$-$\Omega$ microwave drive line is performed by shorting the end of a CPW and splitting the current in the center trace into two paths (shunting wire shown in orange).  The offset $\delta$ from the center of the coils where the wire path splits in two results in asymmetric coupling to the two coils, allowing for an adjustable amount of coupling between the even and odd supermodes of the coupled LC resonators.  Parameters for this device are $d=12~\mu$m and $\delta=16.5~\mu$m, resulting in a simulated coil coupling of $2J/2\pi=415$~MHz and external coupling rates of $\kappaeOdd/2\pi=9.6$~MHz ($\kappaeEven/2\pi=102$~kHz) for the odd (even) mode. The bare (decoupled) frequency of the LC resonators is designed to be $\omegac/2\pi=10.86$ GHz for a capacitor vacuum gap of $s=70$~nm. \textbf{b}, Zoomed-in SEM image of the nanobeam phononic crystal region of the acoustic cavity coupled to coil 2. The Si device layer and Al electrodes appear dark and light grey, respectively. A mirror cell of the surrounding acoustic bandgap shield is shaded in blue.  Inset: SEM of a unit cell of the nanobeam phononic cavity indicating the $s=70$~nm gap size at room temperature.}\label{fig:fig3}
\end{center}
\end{figure*}

A final design consideration relates to the large hypersound frequency of the breathing mode.  Coupling this mode to a microwave circuit of comparable frequency introduces a large effective detuning in the parametric interaction, greatly increasing the required microwave pump power.  We circumvent this problem by using a multimode microwave cavity~\cite{Dobrindt2010} consisting of two coupled single-mode LC resonators. In this scheme, one of the electromagnetic modes is resonant with the microwave pump tone, while the second mode is detuned by the acoustic mode frequency. Formally, the coupling of the acoustic mode to the multimode microwave cavity can be described via the Hamiltonian

\begin{align}
\hat{H} = &\hbar\omegac (\adag_1\ahat_1+\adag_2\ahat_2) +\hbar J(\ahat_1\adag_2+\adag_1\ahat_2)  \nonumber\\ 
&+\hbar\omegam\bdag\bhat  +\hbar g_0 \adag_1\ahat_1(\bhat+\bdag),
\label{eq:H}
\end{align}

\noindent where $J$ is the photon tunneling rate between the two local microwave cavity modes, $\ahat_j (\adag_j)$ are the annihilation (creation) photon operators of local microwave cavity mode $j$, and $\bhat$ ($\bdag$) is the annihilation (creation) operator of the mechanical mode of the acoustic cavity. Diagonalizing this Hamiltonian in a basis of even and odd superpositions of the local microwave cavity modes, and linearizing the interaction for the special case of a strong red-sideband microwave pump field~\cite{Aspelmeyer2014} yields $\Hem'=\hbar \GOM(\ahat_+\bdag+\adag_+\bhat)$,  where $\GOM$ is the parametrically enhanced electromechanical coupling rate, $\GOM= \gzeroEeff \sqrt{\ndriveOdd}$, $\gzeroEeff = \gzeroE/2$, and $\ndriveOdd$ is the intra-cavity photon number due to the strong pump.  In this scenario efficient microwave pumping is realized for $2J \approx \omegam$, corresponding to two-photon resonance, and the microwave pump is at a drive frequency of $\omegad = \omegacEven- \omegam \approx \omegacOdd$, where $\omegacEven$ ($\omegacOdd$) is the frequency of the upper even-symmetry (lower odd-symmetry) supermode of the microwave cavity.  The resulting electromechanical back-action scattering rate between microwave photons and acoustic phonons is $\gammaem = 4\GOM^2/\kappaEven$, where $\kappaEven$ is the decay rate of the even-symmetry cavity mode (see App.~\ref{app:C} for full derivation).  

Figure~\ref{fig:fig3}(a) shows a scanning electron microscope (SEM) image of a fabricated version of the double-cavity device.  This device is fabricated using an Al-on-SOI process introduced in Ref.~\cite{Dieterle2016}, and consists of two LC lumped element microwave resonators which are inductively coupled to each other and capacitively coupled to a pair of hypersonic phononic crystal cavities of slightly different ($5$~MHz) design frequency.  The Al layer is deposited using electron-beam evaporation, and patterning of the Si membrane and Al wiring is performed using electron beam lithography and a combination of plasma dry etching and lift-off.  A SOI wafer with high resistivity ($\gtrsim 5$~k$\Omega$) Si device and handle layers is used to reduce the microwave losses, and the buried-oxide (BOX) layer underlying the Si device layer is removed using an anhydrous hydroflouric acid vapor etch to avoid Al etching and stiction during the membrane release.  Removal of the BOX layer is performed over an extended region covering the entire double-cavity structure in order to avoid the significant microwave losses in the BOX layer.  Figure~\ref{fig:fig3} shows a zoomed-in SEM image of one of the nanobeam acoustic cavities, indicating the placement of the Al capacitor electrodes and the approximately $70$~nm vacuum-gap measured at room temperature.             

\begin{figure}[t!]
\begin{center}
\includegraphics[width=\columnwidth]{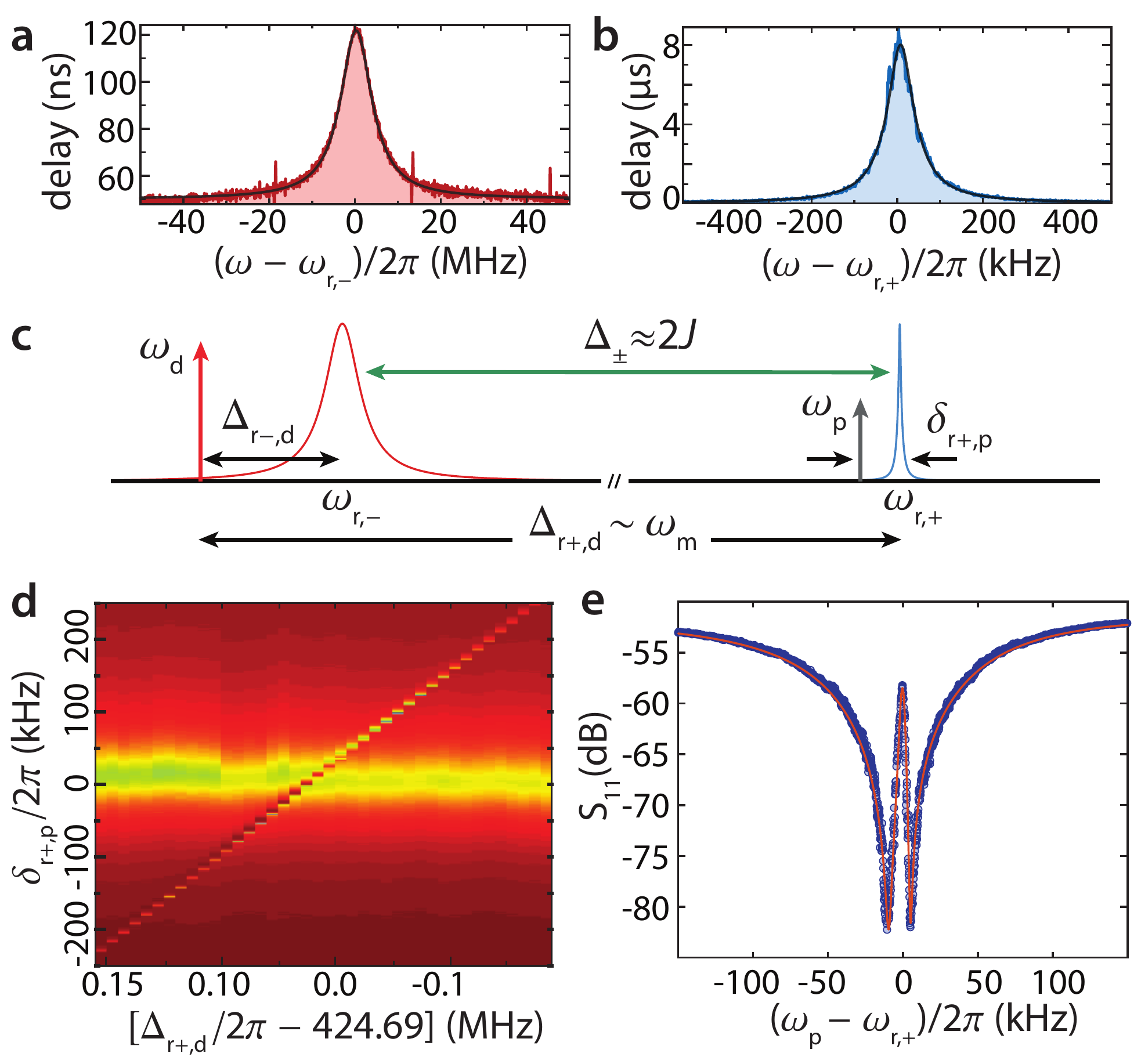}
\caption{\textbf{Microwave electromechanical spectroscopy.} \textbf{a}, VNA measurement of the time delay response of the lower frequency, odd-symmetry-like microwave resonance centered at $\omegacOdd/2\pi = 10.5788$~MHz. \textbf{b}, Delay measurement of the higher frequency, even-symmetry-like resonance centered at $\omegacEven/2\pi = 10.9930$~MHz.  \textbf{c}, Schematic showing the drive and probe frequencies and relevant LC resonance detunings used in the two-tone microwave spectroscopy measurements. Here, $\omegad$ is the drive frequency, which is placed near the low frequency LC resonance ($\omegacOdd$) and close to resonance with the lower motional sideband (red sideband) of the high frequency LC resonance ($\omegacEven$). The high frequency cavity response is registered by sweeping the frequency ($\omegap$) of the weak probe tone. \textbf{d}, Color plot of the measured normalized reflection spectrum, $|S_{11}|$, of a weak coherent probe tone as a function of the detuning ($\DeltardEven$) of the strong drive tone. \textbf{e}, Plot of the measured probe tone spectrum around the high frequency LC resonance for a drive detuning at two-photon resonance, $\DeltardEven \approx \omegam = 424.7$~MHz.  In (\textbf{d}) and (\textbf{e}) the strength of the drive tone corresponds to an intra-cavity photon number in the lower LC anti-symmetric resonance of $\ndriveOdd = 2.25 \times 10^5$.   
}\label{fig:fig4}
\end{center}
\end{figure}

\vspace{2mm}\noindent\textbf{Electromechanical coupling and acoustic damping}

\noindent Measurements of the EMC are performed in a dilution refrigerator with a base temperature of $\Tf=10$ mK (see App.~\ref{app:D}). As detailed in the caption of Fig.~\ref{fig:fig3}, microwave signals are launched onto a $50$-$\Omega$ coplanar waveguide (CPW) on the SOI chip, the center conductor of which is extended and shorted to the ground in the near-field of the cavity. Electrical excitation of the circuit is provided by inductive coupling to the two spiral inductors, which for asymmetric placement of the shorting wire (see Fig.~\ref{fig:fig3}(a)) can lead to stronger loading of the anti-symmetric, lower frequency supermode which is used for pumping of the circuit.  Read-out is performed in reflection.  Vector network analyzer (VNA) microwave delay measurements of the device in Fig.~\ref{fig:fig3} are plotted in Figs.~\ref{fig:fig4}(a) and \ref{fig:fig4}(b), showing the presence of a heavily loaded low-$Q$ resonance at $\omegacOdd/2\pi=10.5788$~GHz and a more weakly coupled high-$Q$ resonance at $\omegacEven/2\pi=10.9930$~GHz.  The measured splitting of $\Delta_{\pm}/2\pi = 414.2$~MHz is in close correspondence to the design tunnel-coupling rate of $2J/2\pi = 415$~MHz.  Fitting the measured delay curves to a Lorentzian lineshape, we infer a total damping rate of $\kappaEven/2\pi=230$~kHz and an external coupling rate of $\kappaeEven/2\pi=85$~kHz for the microwave mode at $\omegacEven$.  For the more strongly loaded resonance at $\omegacOdd$ we measure $\kappaOdd/2\pi \approx \kappaeOdd/2\pi = 8.9$~MHz.  The asymmetry and magnitude of the external coupling of both resonances are again close to the design values for the even and odd modes of the two coil resonator of Fig.~\ref{fig:fig3}. Although the measured splitting and external coil coupling indicate that the local modes of this device are strongly hybridized,  more definitive evidence is provided by the measurement of the cross-mode electromechanical coupling $\gzeroEeff$.  

In order to probe the acoustic properties of the EMC we use a two-tone pump and probe scheme~\cite{Weis2010,Safavi-Naeini2011,Teufel2011a} illustrated schematically in Fig.~\ref{fig:fig4}(c). In this scheme a strong drive tone ($\omegad$) is applied at a variable detuning from the lower frequency microwave resonance while a weaker probe tone ($\omegap$) is scanned across the upper frequency microwave resonance.  At two-photon resonance when the pump-probe difference frequency matches that of the capacitively-coupled acoustic mode frequency, $\omegap-\omegad = \omegam$, beating of the drive and probe tones inside the microwave cavity coherently drive the acoustic mode, leading to interference effects in the externally detected probe tone spectrum due to emission of motional sidebands of the intra-cavity drive field. This results in the emergence of a transparency window in the reflected probe spectrum due to the `dressed' acoustic mode, similar to the phenomena of electromagnetically-induced transparency in atomic physics~\cite{Safavi-Naeini2011}. In the weak-coupling, sideband-resolved limit the reflected probe spectrum is given by:

\begin{equation}
S_{11}=1- \frac{\kappaeEven}{\frac{\kappaEven}{2}+i \deltarpEven+\frac{2\GOM^2}{\gammai+2i(\deltarpEven-(\omegam-\DeltardEven))}},
\label{eq:EIT}
\end{equation}

\noindent where $\deltarpEven \equiv \omegap-\omegacEven$ and $\DeltardEven \equiv \omegacEven-\omegad \approx \omegam$.  Figure~\ref{fig:fig4}(d) shows a color plot of a series of probe scans as the pump detuning is stepped in frequency, showing the emergence of the acoustic resonance near $\DeltardEven/2\pi = 424.69$~MHz.  A plot of the reflected probe spectrum for $\DeltardEven/2\pi = \omegam/2\pi = 424.7$~MHz and $\ndriveOdd = 2.25 \times 10^5$ photons is shown in Fig.~\ref{fig:fig4}(e).  Fitting Eq.~\ref{eq:EIT} to the measured probe spectrum we find a total linewidth of $\gamma/2\pi=6.8$~kHz for the acoustic mode.  This linewidth contains contributions from electromechanical back-action ($\gammaem$), intrinsic energy damping of the acoustic mode ($\gammai$), and any pure dephasing of the acoustic mode.  

\begin{figure}[t!]
\begin{center}
\includegraphics[width=\columnwidth]{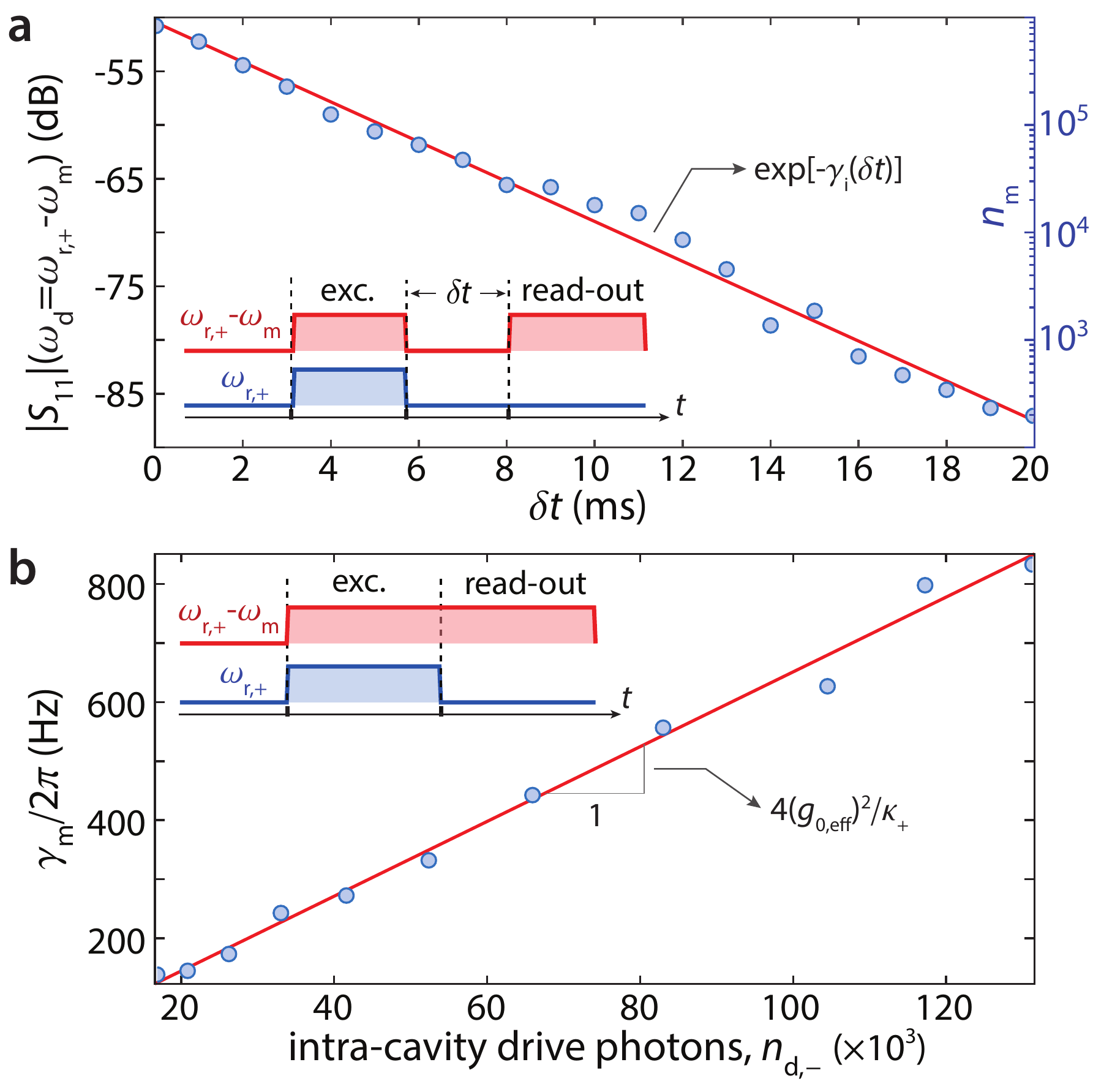}
\caption{\textbf{Mechanical ringdown and electromechanical coupling.} \textbf{a}, Pulsed excitation and read-out measurement of the ringdown of the acoustic energy in the nanobeam resonator.  Here ringdown occurs `in the dark' with all microwave fields off.  \textbf{b}, The measured energy decay rate of the acoustic resonator, $\gamma$, as a function of the power of the drive tone. Here the drive tone is kept on after excitation and the energy decay includes parametric back-action (and parasitic) damping due to the drive tone (see App.~\ref{app:E}).  The vacuum electromechanical coupling $\gzeroE$ is extracted by fitting the measured decay rate $\gamma$ versus the estimated intra-cavity photon number of the drive tone ($\ndriveOdd$).
}\label{fig:fig5}
\end{center}
\end{figure}

We implement a time-domain measurement to determine directly the intrinsic acoustic damping and the electromechanical back-action.  As in the EIT-like spectroscopy, here we apply a $100$~ms two-tone pulse to ring up the mechanics with a strong drive tone at $\omegad = \omegacEven - \omegam$ and a weak probe tone at $\omegap = \omegacEven$.  Detection of the acoustic mode occupancy is performed with a read-out pulse in which the weak probe tone is turned off and motionally scattered photons from the strong drive tone (still at $\omegad = \omegacEven - \omegam$) are detected on a spectrum analyzer in zero-span mode with center frequency at $\omegacEven$ and resolution bandwidth (RBW) set to $30$~kHz ($\gg \gamma/2\pi$).  Figure~\ref{fig:fig5}(a) shows a plot of the measured ringdown of the acoustic mode energy as a function of the delay, $\delta t$, between the end of the ring up pulse and the beginning of the read-out pulse (see inset).  This `ringdown in the dark' measurement yields the intrinsic energy damping rate of the mechanics, which for the breathing mode at $\omegam/2\pi = 424.7$~MHz is measured to be $\gammai/2\pi=68$~Hz (phonon lifetime $\tau=2.3$~ms), corresponding to a $Q$-factor of $6.25 \times 10^6$.  

To extract the back-action induced damping rate of the breathing mode, the read-out pulse delay is set to $\delta t = 0$ and the motionally scattered photons within the read-out pulse are measured as a function of time.  Fitting an exponential decay curve to the time-varying detected read-out signal on the spectrum analyzer for varying read-out pulse powers, Fig.~\ref{fig:fig5}(b) plots the back-action damped acoustic energy decay rate, $\gammam=\gammai+\gammaem$, versus pulse amplitude in units of intra-cavity photon number, $\ndriveOdd$.  As the back-action damping rate for red-sideband pumping is given by $\gammaem \approx 4\gzeroEeff^2\ndriveOdd/\kappaEven$, the slope of this plot yields a vacuum electromechanical coupling rate for the breathing mode of $\gzeroEeff/2\pi = 17.3$~Hz (corresponding to $\gzeroE/2\pi=34.6$~Hz).  Referring to Fig.~\ref{fig:fig1}(d), this value is substantially larger than that expected for the vacuum-gap size of $s=70$~nm measured via SEM at room temperature.  We attribute this difference to a shrinking of the gap to $s\approx 40$~nm due to an increase in the tensile strain of the Al wires on the nanobeam as the device is cooled to cryogenic temperatures.  This is consistent with the observation that devices with gaps smaller than $s \lesssim 60$~nm at room temperature did not show any acoustic resonances when cooled down (the lack of a second acoustic resonance in the device studied here being an example).   



\vspace{2mm}\noindent\textbf{Frequency jitter and thermal occupancy}

\noindent As quantified by the cooperativity, $C \equiv \gammaem/\gammai$, and evidenced in the two-tone spectroscopy measurements of Fig.~\ref{fig:fig3}, coherent manipulation of the breathing mode via microwave drive fields is possible.  A maximum cooperativity of $C \approx 30$ is obtained in this work at the highest attainable drive power of $\ndriveEven = 4.3 \times 10^5$ (see App.~\ref{app:F}). For quantum applications, however, one is also interested in the quantum cooperativity defined as $\Ceff \equiv C/\nbath$, where $\nbath$ is a noise occupancy (Bose factor) of the bath coupled to the acoustic mode.  $\Ceff > 1$ allows for coherent manipulations of the mechanics on a timescale faster than decoherence caused by the bath, or in terms of dissipative processes, back-action cooling of the mechanical mode to its quantum ground state.  We explore here the noise baths coupled to the EMC by monitoring the noise power spectrum generated on the upper motional sideband of a pump tone at $\omegad = \omegacEven - \omegam$ (as in the read-out pulse of the ringdown measurements, but with the spectrum analyzer swept over a finite span with narrow RBW).  

\begin{figure*}[t!]
\begin{center}
\includegraphics[width=\textwidth]{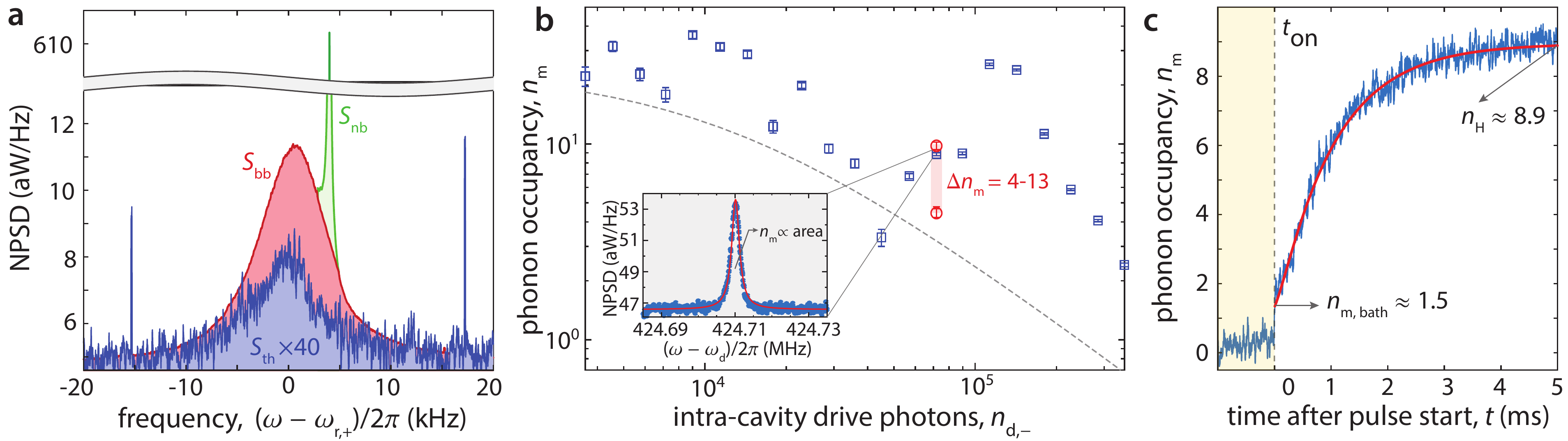}
\caption{\textbf{Frequency jitter noise and mode occupancy.} \textbf{a}, Measured electrical NPSD spectrum at the upper LC resonance ($\omegacEven$) under red-detuned pumping with drive frequency $\DeltardEven \approx \omegam$ and $\ndriveOdd =4.3 \times 10^5$, with and without an additional weak tone at $\omegacEven/2\pi + 4$~kHz. With the weak tone off, the measured NPSD is just that of the back-action damped acoustic resonance ($S_{\text{th}}$; blue curve, magnified by $\times 40$).  With the weak tone on, the measured noise spectrum can be separated into two distinct components, a broad thermal-like spectrum $S_{\text{bb}}$ (red curve) and a narrow noise peak around the weak coherent tone $S_{\text{nb}}$ (green curve).  \textbf{b}, Back-action cooling curve showing measured phonon occupancy versus drive tone power for a pump frequency $\DeltardEven \approx \omegam$.  Dashed grey curve corresponds to expected back-action cooling curve for bath temperature $\Tbath = 500$~mK, $\gammai = 68$~Hz.  $\ndriveOdd = 1$ corresponds to $P_{\text{in}} = 0.62$~fW at the input to the circuit.  Inset: measured NPSD at a drive power of $\ndriveOdd = 7.1 \times 10^4$.  \textbf{c}, Calibrated noise power in units of breathing mode phonon number, $\nm$, versus time during a red-detuned ($\omegad=\omegacEven-\omegam$) drive pulse.  Here a pulse train with pulse period $T_\text{per}=30$~ms and on-pulse length $T_\text{on}=15$~ms is utilized. Pulse amplitude corresponds to an intra-cavity drive photon number $\ndriveOdd = 7.1 \times 10^4$.  The solid red curve shows a heating model fit to the data (see App.~\ref{app:H}) with pump-induced hot bath occupancy $n_H=8.9$.  
}\label{fig:fig6}
\end{center}
\end{figure*}

Plotted in Fig.~\ref{fig:fig6}(a) is the noise power spectral density (NPSD) measured at the upper motional sideband of the pump tone ($\approx \omegacEven$).  In this plot the blue shaded spectrum labelled $S_{\text{th}}$ is the back-action damped noise spectrum of the breathing mode, magnified by a factor of $\times 40$ for visibility.  The measured linewidth of this spectrum is $6.7$~kHz, substantially larger than the estimated linewidth from back-action and intrinsic damping alone ($[\gammai + \gammaem]/2\pi \approx 2$~kHz).  This broadening of the mechanical spectrum is due to time-averaging of the aforementioned frequency jitter.  Although a microscopic understanding of the source of the measured frequency jitter is beyond the scope of this work, an estimate of the time scale of the jitter noise can be determined using the method described in Refs.~\cite{Zhang2014,Zhang2015} in which a weak coherent tone is applied close to the upper motional sideband.  The resulting NPSD is shown in Fig.~\ref{fig:fig6}(a), where the spectrum is separated into a broad thermal-like spectrum ($S_{\text{bb}}$; red curve) and a narrower spectrum around the applied coherent tone ($S_{\text{nb}}$; green curve). Acoustic frequency fluctuations faster than the instantaneous linewidth (given by $\gammai+\gammaem$) contribute to $S_{\text{bb}}$ whereas slower frequency fluctuations add to the narrow spectrum $S_{\text{nb}}$, from which we estimate that $15\%$ ($58\%$) of the measured total linewidth is a result of fast (slow) frequency jitter noise (see App.~\ref{app:G} for details).

Calibration of the reflected signal amplification (along with the electromechanical back-action rate) allows us to relate the area under the NPSD spectrum $S_{\text{th}}$ to an estimate of the noise phonon occupancy of the breathing mode, $\nm$~\cite{Aspelmeyer2014}.  A plot of the inferred $\nm$ versus drive photon number $\ndriveOdd$ is shown in Fig.~\ref{fig:fig6}(b).  Although significant back-action cooling~\cite{Marquardt2007} is expected based upon the measured cooperativity, the variation in $\nm$ versus drive power is highly irregular.  The breathing mode starts out hot at low power, following a weak cooling trend with large fluctuations between different drive powers.  Repeated measurement of $\nm$ at a single drive power also show fluctuations (see red data points at $\ndriveOdd = 7.1 \times 10^4$ in Fig.~\ref{fig:fig6}(b)).  In order to confirm that the breathing mode is thermalized close to the fridge temperature in absence of microwave driving of the circuit, we plot in Fig.~\ref{fig:fig6}(c) the total measured acoustic noise power versus time as the red-sideband cooling pump tone is pulsed on.  The calibrated breathing mode occupancy at the onset of the pulse is measured to be $\nm = 1.5$ (mode temperature $\Tm \approx 40$~mK), and then heats over several milliseconds up to an occupancy $n_H=8.9$.  Over a train of pulses we observe the value of $n_H$ to fluctuate on timescales of a few seconds to minutes, with a variance consistent with the continuous-wave mode occupancy measurements.

The source(s) of the frequency jitter and anomalous mechanical heating observed in current devices is not well understood at this point; however, there are a few candidate sources to consider.  Two-level tunneling systems (TLS)~\cite{Phillips1972} found within amorphous surface oxide layers are known to cause excess damping and noise in microwave superconducting quantum circuits~\cite{Gao2008}. TLS also couple to strain fields, and the high frequency of the hypersonic breathing mode may lead to coupling (via nonlinear phonon-phonon or TLS-phonon scattering) to the same TLS bath as that of the microwave pump.  The observed fluctuation in the breathing mode occupancy is also reminiscent of the bursty nature of quasi-particle (QP) generation measured in thin films of superconducting Al~\cite{Visser2012,Gruenhaupt2018} due to high energy particle impacts.  Modelling of the interaction of QPs with sub-gap electromagnetic radiation indicates that even weak microwave probe fields can lead to non-equilibrium QP and phonon distributions above that of the thermal background~\cite{Goldie2013}.  Both TLS and QP considerations indicate that moving to a different superconducting material with short QP relaxation time and a clean surface -- such as NbTiN~\cite{Barends2010a} -- may significantly reduce acoustic mode heating and frequency jitter.  An additional attribute of NbTiN is its large kinetic inductance~\cite{Samkharadze2016}, which can be employed to further reduce parasitic capacitance and increase the electromechanical coupling.

\vspace{2mm}\noindent\textbf{Conclusions}

\noindent The EMC device presented in this work involves a multimode superconducting microwave circuit coupled to a hypersonic phononic crystal acoustic cavity.  Fabrication is performed using an Al-on-SOI materials platform compatible with both silicon photonics and superconducting qubits~\cite{Dieterle2016,Keller2017}. The design of the acoustic cavity utilizes a phononic crystal to enable efficient coupling to a single mode at frequency $\omegam/2\pi=0.425$~GHz.  Use of phononic bandgap shielding further reduces the coupling of the acoustic mode to external modes of the substrate, resulting in a measured intrinsic $Q$-factor of $6.25 \times 10^6$ for the localized acoustic resonance.  A cooperative coupling between the microwave electrical circuit and the hypersonic acoustic mode of $C\approx 30$ is achieved, enabling sensitive electrical measurement of the acoustic mode close to its quantum ground-state (thermalized mode occupancy $\nm = 1.5$).  

One of the primary motivations of the EMC device development is its application as a quantum transducer, either between microwave electrical and acoustic signals~\cite{Bochmann2013,Arrangoiz_Arriola2018} or between microwave and optical photons when combined with optomechanical upconversion~\cite{Safavi2011,Andrews2014,Bochmann2013,Balram2016}.  In such an application, an EMC transducer may be used to interconnect multiple elements via phononic bandgap waveguides as has been recently demonstrated in optomechanical systems~\cite{Balram2016,Fang2016,Patel2018}.  For chip-scale integrated converters the hypersonic acoustic frequency of the EMC should enable the resolved sideband limit on the optical end, critical for low-noise operation~\cite{Hill2012}, and offers the prospect of high bandwidth transduction in the MHz range, desirable for many applications involving quantum information processing with superconducting qubits~\cite{Devoret2013}.  Improvements in circuit design can help further increase the back-action rate, operational bandwidth, and the acoustic frequency of the EMC studied here. One approach would be to use a lower frequency drive or a DC source~\cite{Pirkkalainen2013,Rouxinol2016} as the parametric pump (effectively replacing the lower frequency microwave resonance in the current devices), which could dramatically reduce the heating caused by absorption of drive photons. A simple scaling of the device used here indicates that for an applied bias of $10$~V, electromechanical coupling at the MHz-level should be possible to acoustic modes at frequencies up to several GHz.     




\begin{acknowledgments}
This work was supported by the AFOSR MURI Wiring Quantum Networks with Mechanical Transducers (grant FA9550-15-1-0015), the ARO-LPS Cross-Quantum Technology Systems program (grant W911NF-18-1-0103), the Institute for Quantum Information and Matter, an NSF Physics Frontiers Center (grant PHY-1125565) with support of the Gordon and Betty Moore Foundation, and the Kavli Nanoscience Institute at Caltech.  M.M. (J.M.F.) gratefully acknowledges support from a KNI (IQIM) Postdoctoral Fellowship.
\end{acknowledgments}


%

\appendix
\clearpage

\onecolumngrid

\section{Circuit properties and coil design}
\label{app:A}

\begin{figure}[h]
\begin{center}
\includegraphics[width=0.8 \columnwidth]{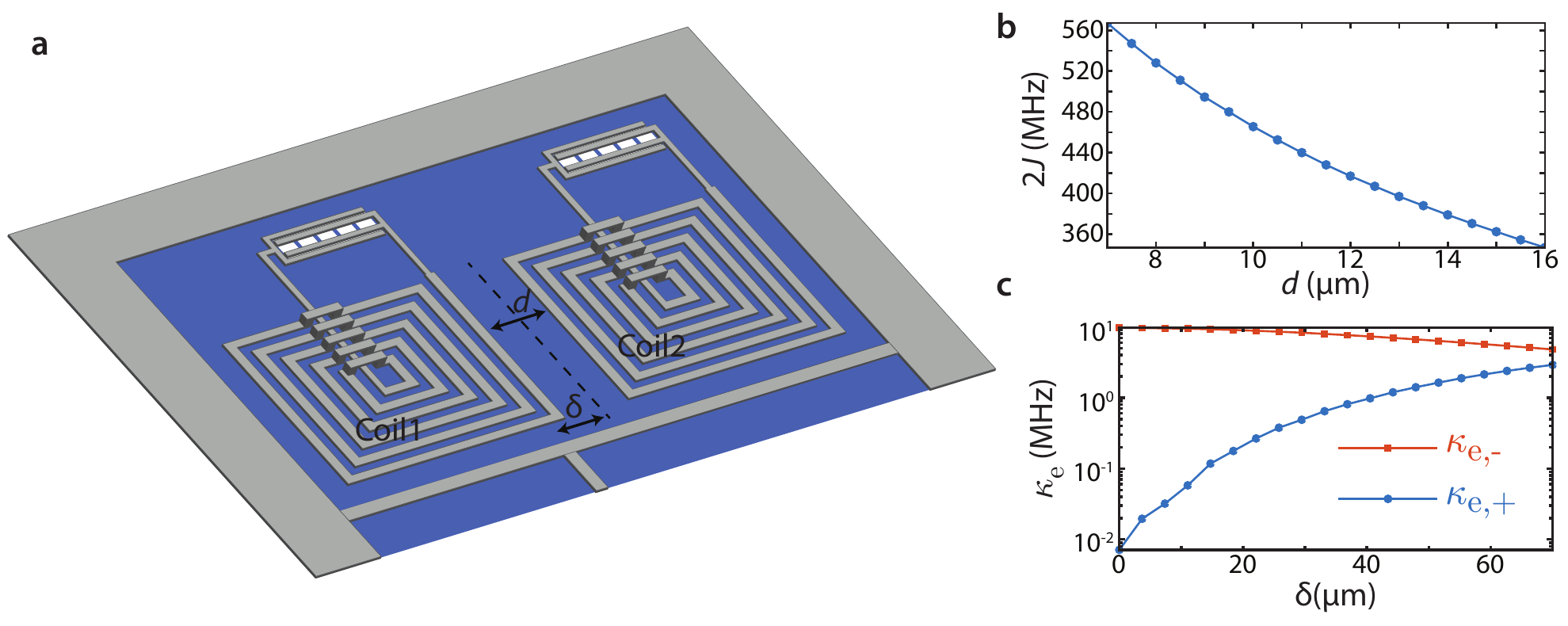}
\caption{\textbf{Planar spiral coil inductor and lumped element LC resonators.}  \textbf{a,} A schematic of the full electromechanical circuit. The microwave circuit is made of aluminum shown in gray color and the silicon membrane is shown in blue. The two LC microwave resonators consist of a high impedance coil which is capacitively coupled to a high frequency mechanical beam. Moreover, the microwave resonators are inductively coupled to each other. Tuning the distance between the two coils $d$ allows for control over the strength of the coupling between the two LC resonators. Additionally, positioning of the two LC resonator system with respect to the branching point of the coupler wire allows for fine tuning of coupling of the even and odd supermodes to the microwave feed line. \textbf{b}, Shows the frequency splitting of the even and odd super modes $2J=\omegacEven-\omegacOdd$ versus the distance between the coils. \textbf{c,} External coupling of even (odd) $\kappaeEven (\kappaeOdd)$ to the microwave feed line versus the separation between branching point of the coupler wire and the center of the two LC resonators $\delta$.}
\label{fig:MicrowaveCircuit}
\end{center}
\end{figure}

The electromechanical circuit is composed of two identical LC coil resonators, each capacitively coupled to a mechanically compliant nanobeam supporting a mechanical mode with slightly different frequencies of 420 MHz and 425 MHz. Fig.~\ref{fig:MicrowaveCircuit}a shows a schematic of the microwave electromechanical circuit. The coil Al wires are 500 nm wide and 120 nm thick, with a 1 $\mu$m pitch and 35 turns forming a square of dimension $74$~$\mu$m$\times 74$~$\mu$m. According to the finite element simulations, the coils have a self resonance frequency at $\omega_{srf}/2\pi = 13.98$~GHz.  This simulation includes the cross-overs and the coupler wires. Using the modified Wheeler formula~\cite{Mohan1999}, the self inductance of the coil is $L= 41.8$~nH.  Using the simple relation $\omega_{srf} =(LC_l)^{-\frac{1}{2}}$, we obtain the total stray capacitance of $C_s = C_c+C_p = 3.1$~fF, where $C_c$ is the parasitic capacitance from the coil inductor and $C_p$ is the extra parasitic capacitance due to the extra wiring. The modulating capacitance $C_m$ for different gap size of the electrodes is shown in Fig.~\ref{fig:fig1}(d) of the main text. From the measured $g_0$ in our experiment we estimate the capacitor gap to be $s\approx 40$~nm, from which we estimate the modulating capacitance to be $C_m = 2.1$~fF. This yields a participation factor of $\eta \approx 0.4$ and a bare LC coil resonance frequency of $\omegac/2\pi = 10.77$~GHz (when adding the estimated motional capacitance to stray capacitance $C_{s}$ of the coil and extra wiring).  Using the simulated value of $2J$, we estimate $\omegacEven = 10.9975$~GHz and $\omegacOdd = 10.5625$~GHz, very close to the measured values of the supermodes which are $10.9930$~GHz and $10.5788$~GHz, respectively. 

The two mechanical beams are designed to have the same length and gap size such that the modulating capacitance and the frequency of both microwave resonators are identical. By placing the two microwave resonator near each other, the two coils are inductively coupled forming an even and odd symmetry microwave supermodes. By adjusting the distance between the two inductors, we can control the strength of the mutual inductance hence, controlling the frequency splitting between the even and odd microwave mode. Figure~\ref{fig:MicrowaveCircuit}b shows the frequency splitting between the two microwave modes as a function of the distance between the coil inductors. For the device presented in this manuscript, the distance is chosen to be $d=12$~$\mu$m yielding a simulated frequency splitting of $2J/2\pi=415$~MHz between the microwave modes. Moreover, the placement of the two inductors with respect the branching point of the coupler wire denoted by $\delta$ allows us to independently control the coupling of the microwave supermodes to the microwave feed line. Fig.~\ref{fig:MicrowaveCircuit}c shows the coupling of the even and odd microwave modes to the microwave feed line versus the displacement of the coil inductors $\delta$. For the chosen $\delta= 16.5$~$\mu$m, we expect the coupling of $\kappaeEven/2\pi= 102$~kHz for the even mode at higher frequency and $\kappaeOdd/2\pi= 9.6$~MHz for the odd mode at lower frequency. All the microwave LC resonator simulations are performed using the SONNET software package~\cite{SONNET}.  In these simulations we have assumed a relative permittivity of $\epsilon_\text{Si} = 11.7$ for the silicon membrane.

\section{Phononic bandgap and mechanical radiation-$Q$ simulation}
\label{app:B}

\begin{figure}[h]
\begin{center}
\includegraphics[width=0.8 \columnwidth]{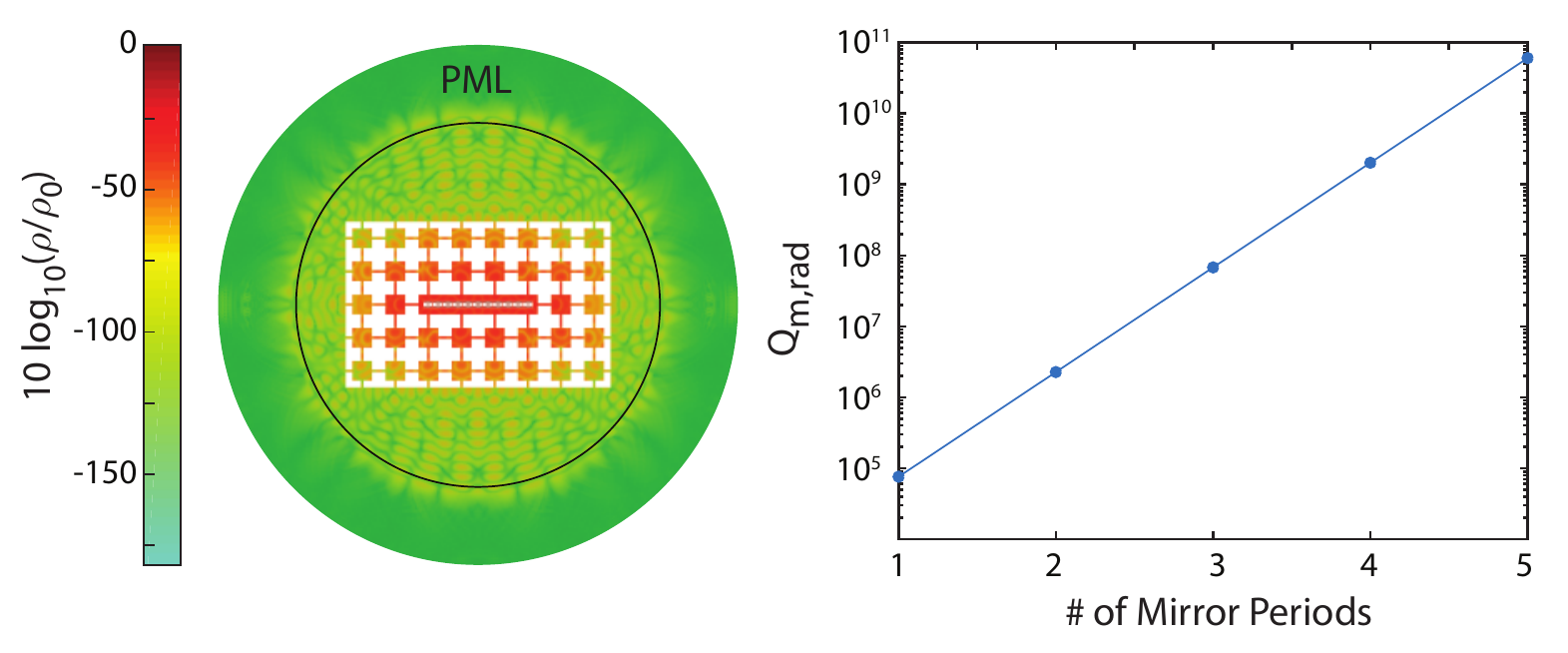}
\caption{\textbf{Mechanical radiation-$Q$ simulations.}  \textbf{a,} FEM Simulation of the mechanical beam resonator surrounded by a 2-period phononic band gap structure, silicon membrane and a perfectly matched layer (PML). Color indicates the normalized mechanical energy density with red (green) color corresponding to the area with highest (lowest) mechanical energy density. \textbf{b,} Simulated radiational mechanical quality factor versus the number of the periods of the phononic band gap shield.}
\label{fig:MechanicalQ}
\end{center}
\end{figure}

Periodic structures can be engineered to have a phononic bandgap where mechanical energy loss by linear elastic coupling to the environment is suppressed. Lack of a phononic bandgap in the nanobeam structure means that the localized mechanical mode of the beam can couple with radiating acoustic modes due to imperfections in the structure that cause acoustic scattering and break the beam symmetry. To prevent the coupling of the localized mechanical modes of the nanobeam with the acoustic modes of the surrounding silicon membrane and substrate we have embedded our mechanical beam resonator in a phononic bandgap `cross' structure (see Fig.~2 in the main text). To model the damping of the localized breathing mode due to acoustic radiation we include a special circular pad region around the structure that absorbs all the outgoing acoustic waves. This domain is called a perfectly matched layer (PML). What distinguishes a PML domain from an ordinary absorbing/lossy material is that all the waves incident upon the PML from a non-PML media do not reflect at the interface, hence the PML strongly absorbs all outgoing waves from the interior of the computational region.

An example of the FEM simulation with PML is shown in Fig.~\ref{fig:MechanicalQ}a. In this image the nanobeam acoustic resonator and a 2-period phononic bandgap shield are surrounded by an inner circular membrane of normal Si material and an outer, larger circular PML domain also made of Si. The color indicates the normalized acoustic energy density of the breathing mode throughout the structure. As shown, with using just 2 periods for the phononic shield, the radiated mechanical energy density is suppressed by $\approx100$~dB. Fig.~\ref{fig:MechanicalQ}b shows the simulated radiation-limited quality factors of the breathing mode of the mechanical resonator versus the number of phononic shield periods surrounding the mechanical nanobeam. All the finite element method simulation of the mechanical nanobeam are done using COMSOL Multiphysics~\cite{COMSOL}. The mechanical simulation use the full anisotropic elasticity matrix where $(C_{11},C_{12},C_{44}) = (166,64,80)$~GPa and assumes a [100] crystalline orientation aligned with the $x$-axis of the nanobeam (direction along its length).

\section{Derivation of two-mode electromechanical response}
\label{app:C}

\begin{figure}[h]
\begin{center}
\includegraphics[width=0.4 \columnwidth]{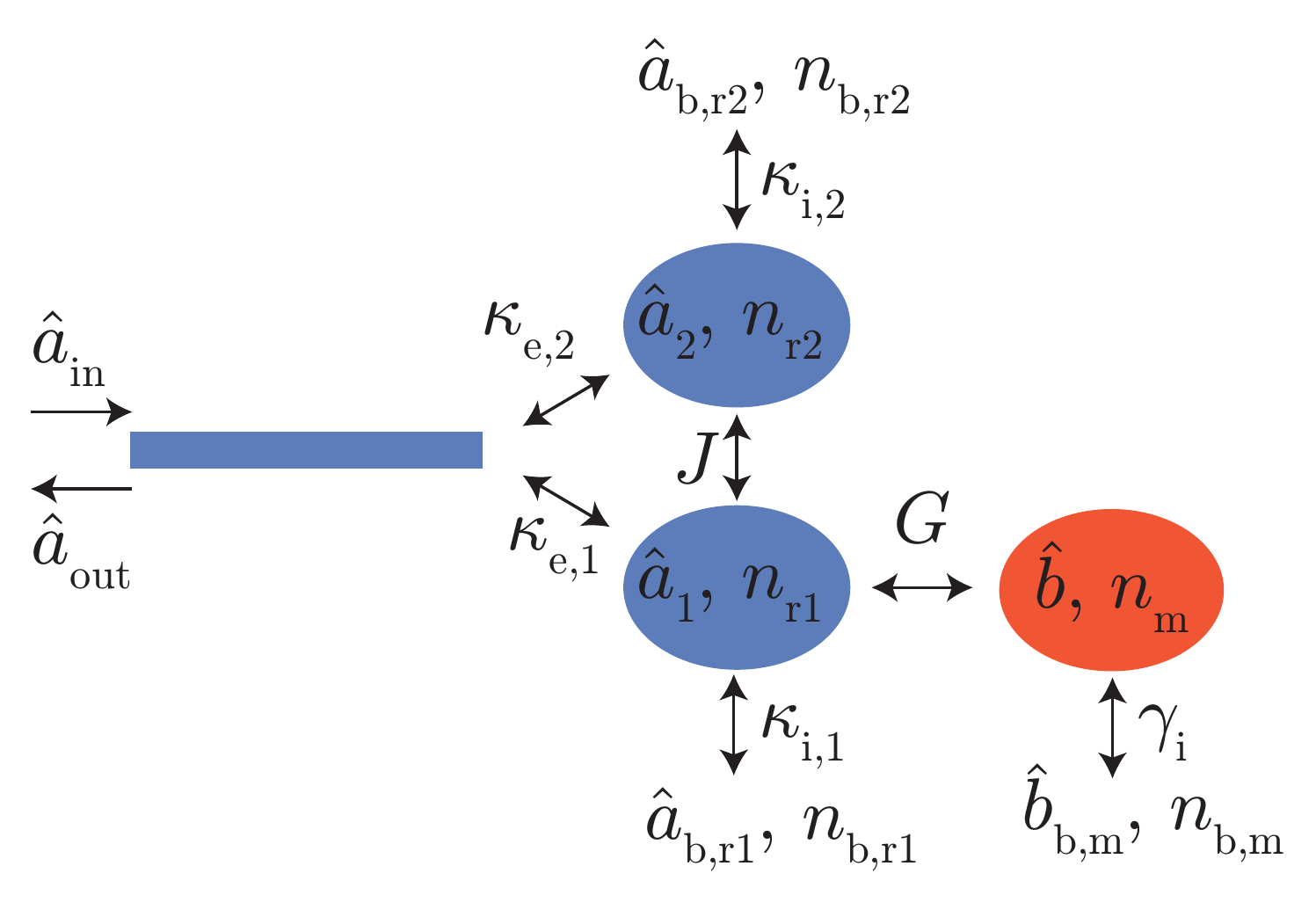}
\caption{\textbf{Input-Output Mode Schematic.} In the reflective geometry the microwave cavity mode $\ahat_{1} (\ahat_2)$ is coupled to the coherent waveguide modes $\ain$ and $\aout$ with the external coupling strength $\kappa_{e,1} (\kappa_{e,2})$. It is also coupled to a bath of noise photons, ideally at the refrigerator temperature $n_{b,r1} (n_{b,r2})$, with the intrinsic coupling strength $\kappa_{i,1} (\kappa_{i,2})$. The mechanical resonator mode $\bhat$ is coupled to the microwave resonator with the parametrically enhanced electromechanical coupling strength $\GOM$. In addition, it is coupled to a bath of noise phonons, ideally at the refrigerator temperature $\nmbath$, with the intrinsic coupling rate $\gammai$.}
\label{fig:HamiltonianSchematic}
\end{center}
\end{figure}

In this section we follow previous work~\cite{Marquardt2007,Dobrindt2008,Teufel2011b,Rocheleau2009,Fink2016,Dobrindt2010} to the calculate coherent response and the noise spectrum of the electromechanical system.

\subsection{System Hamiltonian of two-mode microwave electromechanical system}

The Hamiltonian of the coupled microwave electromechanical system (see Fig.~\ref{fig:HamiltonianSchematic}) can be written as

\begin{equation}
\hat{H}=\hbar \omegac \adag_1\ahat_1 + \hbar\omegac\adag_2\ahat_2+\hbar J (\adag_1\ahat_2+\adag_2\ahat_1)+\hbar \omegam \bdag\bhat+\hbar \gzeroE\adag_1\ahat_1(\bdag+\bhat), 
\end{equation}

\noindent where $\ahat_j(\adag_j)$ are the annihilation (creation) operators of the local microwave cavity modes with bare frequency $\omegac$ and $\bhat (\bdag)$ are the annihilation (creation) operator of the mechanical mode. $J$ is the coupling between the two microwave cavity and $\gzeroE$ is the single photon coupling between the microwave cavity 1 and the mechanical mode. We diagonalize the Hamiltonian by introducing the even and odd superposition of the microwave cavity modes of the form $\ahat_{\pm}=\frac{\ahat1\pm\ahat2}{\sqrt{2}}$.

The new Hamiltonian is now written as 
\begin{equation}
\hat{H'}=\hbar \omegacEven \adag_+\ahat_+ + \hbar\omegacOdd\adag_-\ahat_-+\hbar \omegam \bdag\bhat+\hbar \frac{\gzeroE}{2}(\adag_+\ahat_-+\adag_-\ahat_++\adag_+\ahat_++\adag_-\ahat_-)(\bdag+\bhat),
\end{equation}
where the new supermode frequencies are $\omega_\pm=\omegac \pm J$. We excite the microwave resonators by using a strong coherent drive at frequency $\omegad$ detuned from the microwave resonators frequency by $\Delta_{r\pm,d}=\omega_{r\pm}-\omegad$. Thus we can write the Hamiltonian in the rotating frame

\begin{equation}
\hat{\tilde{H}}=-\hbar \DeltardEven \adag_+\ahat_+ - \hbar\DeltardOdd\adag_-\ahat_-+\hbar \omegam \bdag\bhat+\hbar \frac{\gzeroE}{2}(\adag_+\ahat_-+\adag_-\ahat_++\adag_+\ahat_++\adag_-\ahat_-)(\bdag+\bhat). 
\end{equation}

Assuming a strong red detuned drive from the higher frequency microwave cavity and $\DeltardEven \approx \omegam \gg \DeltardOdd $ we can linearize the Hamiltonian in the rotating frame to obtain
\begin{equation}
\hat{\tilde{H}}=-\hbar \DeltardEven \adag_+\ahat_+ - \hbar\DeltardOdd\adag_-\ahat_-+\hbar \omegam \bdag\bhat+\hbar \GOM(\ahat_-+\adag_-+\adag_++\ahat_+)(\bdag+\bhat),
\label{eq:Hamiltonian} 
\end{equation}
where $\GOM=\frac{\gzeroE}{2}\sqrt{\ndriveOdd}$ and $\ndriveOdd$ corresponds to intra-cavity photon number of the lower frequency microwave cavity defined by
\begin{equation}
\ndriveOdd =\frac{\Pdrive}{\hbar \omegad} \frac{4\kappaeOdd}{\kappaOdd^2 + 4\DeltardOdd^2},
\end{equation}
where $\Pdrive$ is the power at the cavity input, expressed by $\Pdrive= 10^{-3}10^{(\mathcal{A}+P_{in})/10}$ with $P_{in}$ the drive power in dBm and $\mathcal{A}$ the total attenuation of the input line in dB. For the Hamiltonian in Eq.~(\ref{eq:Hamiltonian}), the linearized Langevin equations are given as
\begin{align}
\dot{\ahat}_-(t)  & =  - \left ( i \DeltardOdd + \frac{\kappaOdd}{2} \right ) \ahat_-(t) - i \GOM (\bhat(t) + \bdag(t))-\sqrt{\kappaeOdd} \ain(t) - \sqrt{\kappaiOdd} \abOdd(t), \\
\dot{\ahat}_+(t)  & =  - \left ( i \DeltardEven + \frac{\kappaEven}{2} \right ) \ahat_+(t) - i \GOM (\bhat(t) + \bdag(t))-\sqrt{\kappaeEven} \ain(t) - \sqrt{\kappaiEven} \abEven(t), \\
\dot{\hat{b}}(t) & = - \left ( i \omegam + \frac{\gammai}{2} \right ) - i \GOM (\ahat_+(t)+ \adag_-(t) )- \sqrt{\gammai} \bb(t).
\end{align}
Here, $\hat{a}_{b,r\pm}$ are annihilation operators of the microwave baths coupled to the even and odd microwave cavities. We define the following convention for the Fourier transform. Given an operator $\hat{A}$ we define 
\begin{align}
 \hat{A}(t)&=\frac{1}{\sqrt{2\pi}}\int_{-\infty}^{+\infty} d\omega e^{-i\omega t}\hat{A}(\omega), \\
%
\hat{A}(\omega)&=\frac{1}{\sqrt{2\pi}}\int_{-\infty}^{+\infty} dt e^{i\omega t}\hat{A}(t).
\label{Eq:FourierTransform2}
\end{align}

Now, taking the Fourier transform of the linearized Langevin equations and simplifying we obtain 
\begin{align}
\chirOdd^{-1}(\omega){\ahat_-(\omega)}  & =  - i \GOM (\bhat(\omega) + \bdag(\omega))-\sqrt{\kappaeOdd} \ain(\omega) - \sqrt{\kappaiOdd} \abOdd(\omega), \\
\chirEven^{-1}(\omega){\ahat_+(\omega)}  & =  - i \GOM (\bhat(\omega) + \bdag(\omega))-\sqrt{\kappaeEven} \ain(\omega) - \sqrt{\kappaiEven} \abEven(\omega), \\
\chim^{-1 }(\omega){\bhat(\omega)} &  =  - i \GOM (\ahat_-(\omega) + \adag_-(\omega)+\ahat_+(\omega) + \adag_+(\omega) )-\sqrt{\gammai} \bb(\omega),
\end{align}
where $\chi_{r\pm}$ and $\chim$ are the uncoupled susceptibilities of the electrical and the mechanical modes defined by 
\begin{align}
\chi_{r\pm}^{-1}(\omega) & =\kappa_{\pm}/2 + i ( \Delta_{r\pm,d}-\omega),\\
\chim^{-1}(\omega) & =\gammai/2 + i (\omegam- \omega).
\end{align} 

In the sideband-resolved limit $\omegam \gg \kappa_{\pm},G$ and for positive detuning of the drive tone with respect to the high frequency microwave cavity $\DeltardEven\approx \omegam$(red side pumping), we have $\chim\chirOdd \ll \chim\chirEven$. Thus the linearized Langevin equation can be simplified further and written approximately as 
\begin{align}
\ahat_+(\omega) &=\frac{i \GOM \chirEven\chim \sqrt{\gammai}\bb(\omega) - \chirEven(\sqrt{\kappaeEven}\ain(\omega)+\sqrt{\kappaiEven}\abEven(\omega))}{1+\GOM^2\chirEven\chim}, \label{Eq:linlangEven}\\
\ahat_-(\omega) &= - \chirOdd(\sqrt{\kappaeOdd}\ain(\omega)+\sqrt{\kappaiOdd}\abOdd(\omega)),\label{Eq:linlangOdd}\\
\bhat(\omega) &  =  \frac{-\chim\sqrt{\gammai}\bb(\omega) -i\GOM\chim\chirEven(\sqrt{\kappaeEven}\ain(\omega)+\sqrt{\kappaiEven}\abEven(\omega))}{1+\GOM^2\chirEven\chim},
\label{Eq:linlangMechanics}
\end{align}
where we have dropped the terms proportional $\chim\chirOdd$. For driving of the odd-symmetry microwave mode, transduction of the acoustic mechanical mode is observed in the output field of the even-symmetry microwave cavity mode.  Using the Input-Output formalism and Eq.~(\ref{Eq:linlangEven}) we find for the output field near resonance with the even-symmetry microwave mode
\begin{equation}
\begin{aligned}
\aout(\omega) &= \ain(\omega)+\sqrt{\kappaeEven}\ahat_+(\omega),\\
&= \ain(\omega)-\kappaeEven\frac{\chirEven \ain(\omega)}{1+\GOM^2\chim\chirEven}-\frac{\sqrt{\kappaeEven\kappaiEven}\chirEven\abEven(\omega)}{1+\GOM^2\chim\chirEven}+\bb(\omega)\frac{i\GOM\sqrt{\kappaeEven\gammai}\chim\chirEven}{1+\GOM^2\chim\chirEven}.
\label{Eq:aout}
\end{aligned}
\end{equation}

\subsection{Electromagnetically Induced Transparency}
We first calculate the coherent part of the output signal by using Eq.~(\ref{Eq:aout}) and dropping the incoherent terms to get
\begin{equation}
S_{11}=\frac{\left<\aout(\omega)\right>}{\left<\ain(\omega)\right>}=1- \frac{\kappaeEven\chirEven}{1+\GOM^2\chim\chirEven},
\end{equation}
and substituting for the bare microwave and mechanical cavity susceptibilities we get the coherent electromechanical analogue of the electromagnetic induced transparency valid for small probe powers,
\begin{equation}
S_{11}= 1 - \frac{\kappaeEven}{\kappaEven/2 + i(\Delta_{r+,d}-\omega) +\frac{\GOM^2}{\gammai/2+i(\omegam-\omega)}}.
\end{equation}

\subsection{Quantum derivation of observed noise spectra}
For an operator $\hat{A}$ the power spectral density is written as
\begin{align}
S_{AA}(t)&=\int_{-\infty}^{+\infty} d\tau e^{i\omega\tau}\left<\hat{A}^{\dagger}(t+\tau)\hat{A}(t)\right>.\label{Eq:TimePSD}
\end{align}
Utilizing Fourier transform defined above, we can re-define the power spectral density of an operator $\hat{A}$  in terms of frequency~\cite{Fink2016} as
\begin{align}
S_{AA}(\omega)&=\int_{-\infty}^{+\infty}d\omega'\left<\hat{A}^{\dagger}(\omega)\hat{A}(\omega')\right>.
\end{align}
 Thus, for the auto-correlation of the detected normalized field amplitude (or the photo current) $\hat{I}(t)=\aout(t)+\aout^{\dagger}(t)$, we get
 \begin{equation}
S_{II}=\int_{-\infty}^{+\infty} d\omega'\left<\left(\aout(\omega)+\aout^{\dagger}(\omega)\right)\left(\aout(\omega')+\aout^{\dagger}(\omega')\right)\right>.
\label{Eq:SII}
\end{equation}
Substituting for $\aout(\omega)$ and $\aout^{\dagger}(\omega)$ from Eq.~(\ref{Eq:aout}) and using thermal noise correlation for input noise terms (\emph{i.e.}~$\langle \bb(\omega)\bb^\dagger(\omega')\rangle=(\nmbath+1)\delta(\omega+\omega')$, 
$\langle \bb^\dagger(\omega)\bb(\omega')\rangle=\nmbath\delta(\omega+\omega')$,
$\langle \abEven(\omega)\abEven^\dagger(\omega')\rangle=(\nbEven+1)\delta(\omega+\omega')$,
$\langle \abEven^\dagger(\omega)\abEven(\omega')\rangle=\nbEven\delta(\omega+\omega')$, 
$\langle \ain(\omega)\ain^\dagger(\omega')\rangle=\delta(\omega+\omega')$), the power spectral density is
\begin{multline}
S_{II}(\omega)=
\Big|\Big(1-\frac{\kappaeEven \chirEven}{1+\GOM^2 \chim \chirEven}\Big)\Big|^2
+(\nbEven+1) \frac{\kappaeEven \kappaiEven |\chirEven|^2 }{|1+\GOM^2 \chim \chirEven|^2}
+(\nmbath+1)\frac{\kappaeEven \gammai \GOM^2 |\chim|^2 |\chirEven|^2}{|1+\GOM^2 \chim \chirEven|^2}.
\label{Eq:FullSII}
\end{multline}

\noindent The mechanical occupancy can also be calculated by using Eq.~(\ref{Eq:linlangEven}) and Eq.~(\ref{Eq:linlangMechanics}) as in Ref.~\cite{Dobrindt2008} 
\begin{equation}
\nm=\nmbath \left(\frac{\gammai}{\kappaEven}\frac{4\GOM^2+\kappaEven^2}{4\GOM^2+\kappaEven\gammai}\right) + \nrEven\left( \frac{4\GOM^2}{4\GOM^2 +\kappaEven\gammai}\right).
\label{Eq:nm}
\end{equation}

\section{Experimental Setup}
\label{app:D}

\begin{figure}[h]
\begin{center}
\includegraphics[width=0.65 \columnwidth]{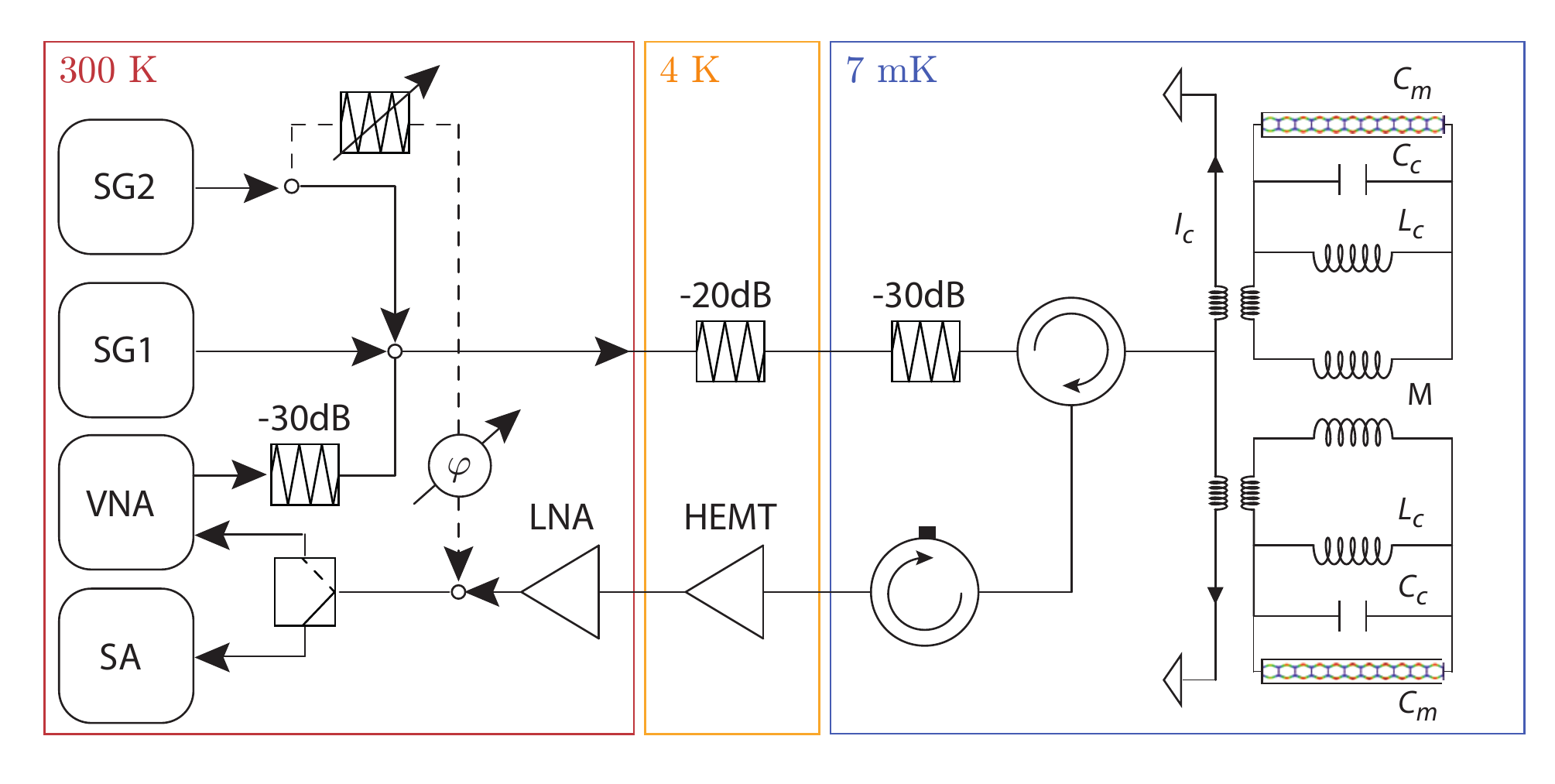}
\caption{\textbf{Experimental setup.}
The output tone of a microwave signal generators (SG1 and SG2) and the output tone of a vector network analyzer (VNA) are combined at room temperature, attenuated, routed to the sample at about 7~mK and inductively coupled to the LC circuit on the silicon membrane. We detect the reflected output tone which is routed and isolated with two microwave circulators and after amplification with a high electron mobility transistor amplifier (HEMT) at 4K stage, switchable pump tone cancelation (dashed lines), and further amplification with a low noise amplifier (LNA). The measurement is done either phase coherently with the VNA, or we detect the incoherent power spectrum with the spectrum analyzer (SA).} 
\label{fig:setup}
\end{center}
\end{figure}

For measuring the coherent and incoherent response of our circuit, we combine the output of a vector network analyzer with two other microwave sources and feed the microwave signals to the base plate of a cryogen free dilution refrigerator at $\Tf\approx 10$~mK using coaxial cables with feedthroughs and attenuators for thermalization at multiple temperature stages yielding total attenuation of $\mathcal{A} = 76$~dB and suppressing room temperature Johnson noise (see Fig.~\ref{fig:setup}) to about $0.05$ photons. We couple to the sample in a reflective geometry using a circulator and a low loss copper printed circuit board (PCB). On the PCB and the chip we use 50~$\Omega$ coplanar waveguides (CPW) to route the microwave tones to the membrane with very little reflections ($<-25~$dB). Near the LC circuit, we extend the center conductor of the CPW waveguide and short it to ground with a narrow wire passing near the inductor coils and inductively couple to the microwave resonators.

On the output side, we use an isolator for isolating the sample from the 4~K stage noise. Niobium titanium superconducting cables are used to connect the isolator directly to a low noise, high electron mobility transistor amplifier (HEMT) at 4~K stage. The microwave signal is again amplified with a low noise room temperature amplifier (LNA). In order to suppress spurious response peaks for high drive power measurements, we cancel the high drive power tone by adding a phase and amplitude adjusted part of the pump tone to the output signal before the room temperature amplifier, as shown in Fig.~\ref{fig:setup}. After the final amplification we use an electronically controlled microwave switch to distribute the signal to either the spectrum analyzer or the second vector network analyzer port. The total gain of the system is $\mathcal{G} = 57.6$~dB.

\section{Mechanical ringdown measurements}
\label{app:E}

\begin{figure}[t!]
\begin{center}
\includegraphics[width=0.7\columnwidth]{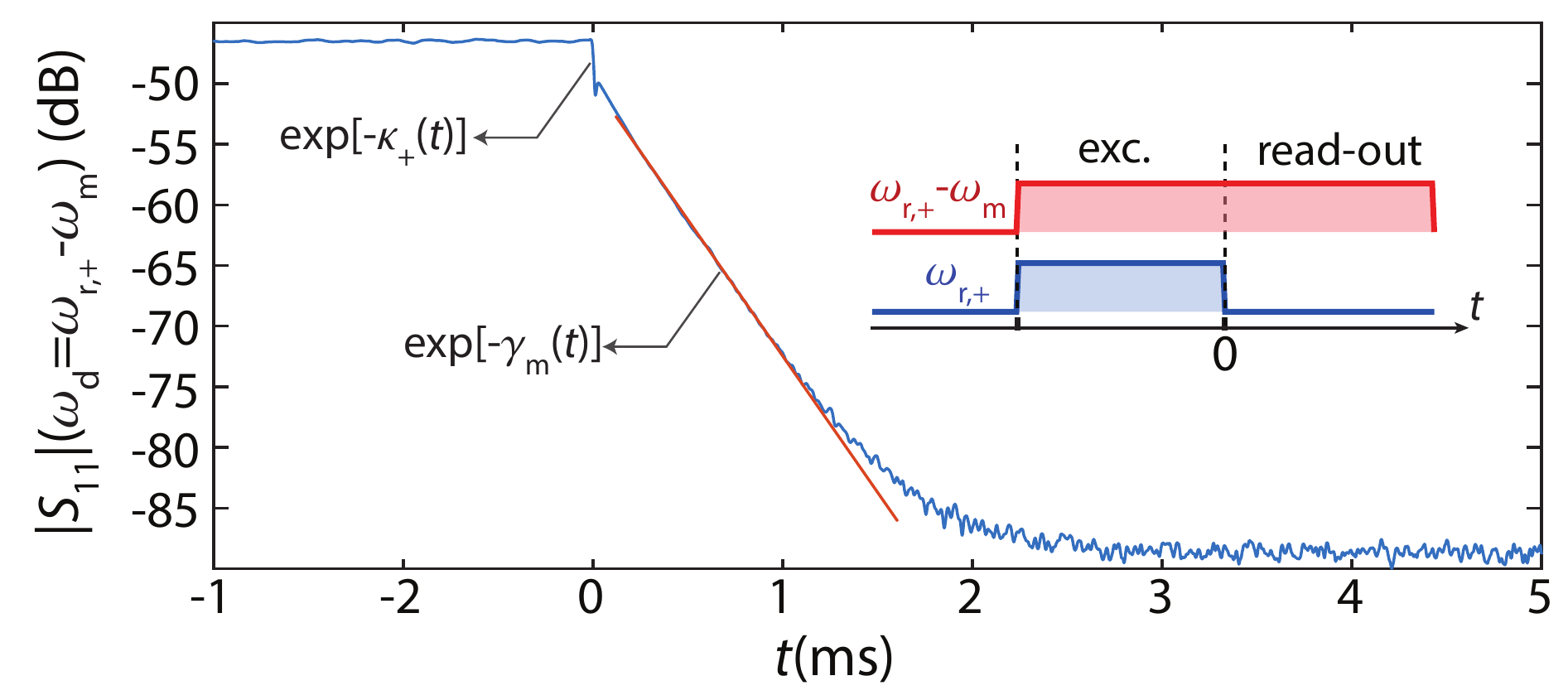}
\caption{\textbf{Mechanical Ringdown Measurement.} Pulsed excitation and read-out measurement of the ringdown of the acoustic energy in the nanobeam mechanical resonator. In this measurement the drive tone at 6 dBm power corresponding to an intra-cavity photon number of $\ndriveOdd = 1.42\times 10^5$ is kept on after the excitation and the decay includes back action damping due to the drive tone. The initial steep decay results from the leakage of the photons from the microwave cavity. The decay rate of the mechanical resonator $\gammam$ is extracted by fitting an exponential curve to the slow decay section of the signal versus time.}\label{fig:ringdown}
\end{center}
\end{figure}

In the ringdown measurements performed in this work we apply, as in the EIT-like spectroscopy, a $100$~ms two-tone pulse to ring up the mechanics with a strong drive tone at $\omegad = \omegacEven - \omegam$ and a weak probe tone at $\omegap = \omegacEven$.  Detection of the acoustic mode occupancy is performed with a read-out pulse in which the weak probe tone is turned off and motionally scattered photons from the strong drive tone (still at $\omegad = \omegacEven - \omegam$) are detected on a spectrum analyzer in zero-span mode with center frequency at $\omegacEven$ and resolution bandwidth (RBW) set to $30$~kHz ($\gg \gamma/2\pi$). A plot of the motionally scattered photons, corresponding the acoustic mode occupancy versus time is shown in Fig.~\ref{fig:ringdown} for $\ndriveOdd = 1.42\times 10^5$ photons. The initial steep decay results from the leakage of the photons from the microwave cavity and is followed by a slower decay of the signal due to the mechanical damping. The measured mechanical damping contains contributions from back-action $\gammaem$ and intrinsic energy damping of the acoustic mode $\gammai$.

\section{Modeling the two-photon response and determining cooperativity}
\label{app:F}

\begin{figure}[h]
\begin{center}
\includegraphics[width= 0.8\textwidth]{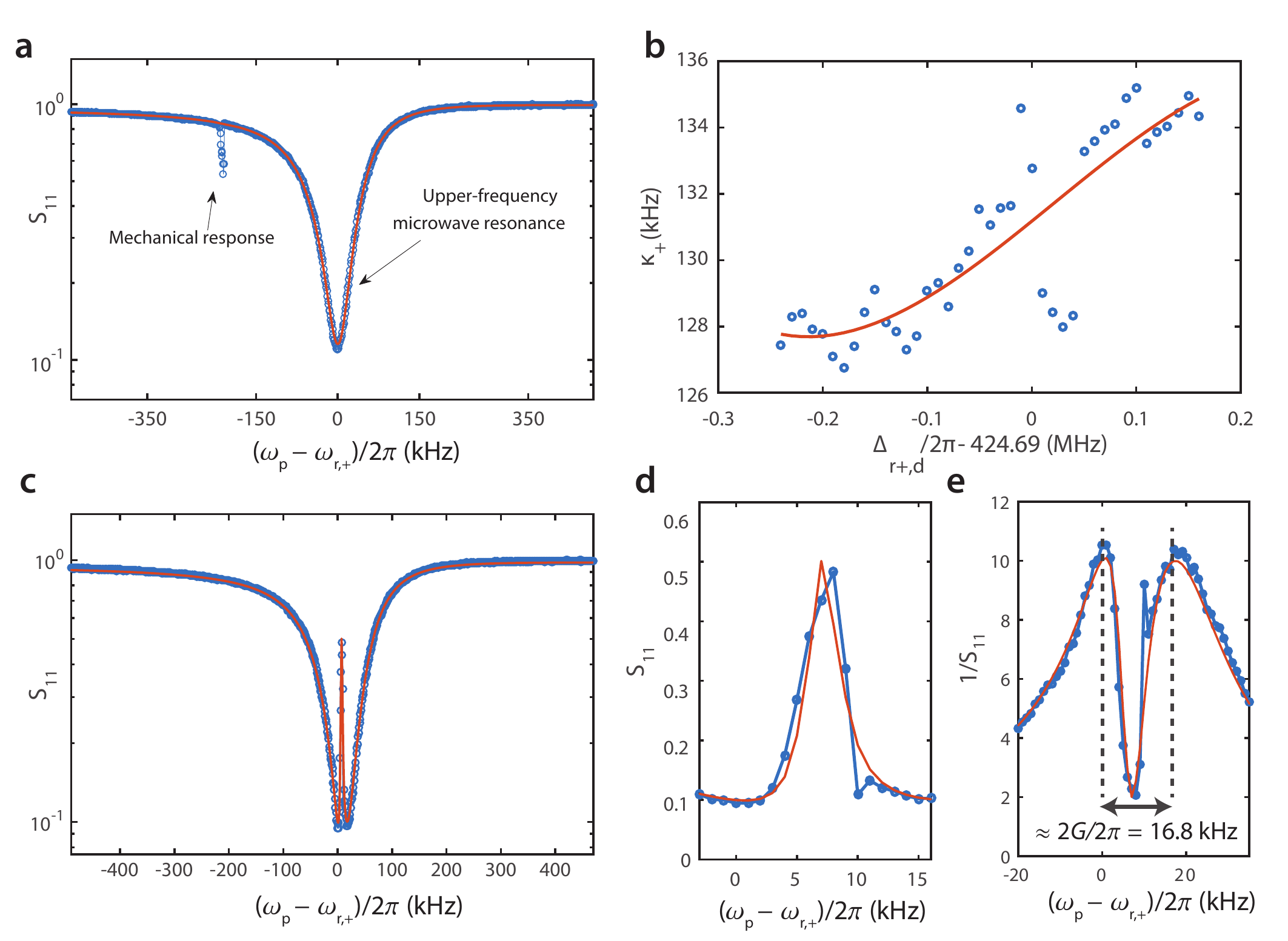}
\caption{\textbf{Modeling of the two-photon response.}  \textbf{a,} Reflection spectrum of a weak coherent probe tone at frequency $\omegap$ close to the upper microwave resonance at $\omegacEven$, in the presence of a strong drive tone at $\omegad$ which is detuned by $\DeltardEven \equiv \omegacEven-\omegad$. The mechanical lineshape and the microwave resonance parameters can be identified separately in this measurement where $\DeltardEven \ne \omegam$. Solid red line shows the theory fit to the microwave resonance assuming a Lorentzian lineshape.  \textbf{b,} The fitted value of total linewidth ($\kappa$) for the upper frequency microwave resonance as a function of the detuning of the strong drive tone ($\DeltardEven$). \textbf{c,} Fitted EIT-like response for zero two-photon detuning ($\DeltardEven \approx \omegam$). \textbf{d,} The central part of the reflection spectrum is used for extracting the jitter in the mechanical frequency. Solid line shows a fit assuming a FWHM jitter of $\delta\omegam/2\pi = 1.98$~kHz. \textbf{e,} Inverse of the probe tone reflection as a function of frequency. The values for parametrically enhanced electromechanical coupling ($\GOM$), and the intrinsic decay rate of the mechanical mode ($\gammai$) are extracted by fitting a theory curve to the two peaks. }
\label{fig:EITfit}
\end{center}
\end{figure}

To probe the acoustic properties of the fabricated device, and to determine the degree of coherent coupling between the microwave and acoustic resonances, we use a two-tone pump and probe scheme.  The presence of acoustic frequency jitter complicates the modeling of the electromechanical response, and so we detail here the methods we employ to determine the electromechanical cooperative coupling, $C$.  In this scheme, described also in the main text, a strong drive tone ($\omegad$) is applied at a variable detuning from the lower frequency microwave resonance while a weaker probe tone ($\omegap$) is scanned across the upper frequency microwave resonance.  Figure~\ref{fig:EITfit}a shows one of the measured spectrum from Fig.~4d of the main text, where the drive power in units of intra-cavity photons is $\ndriveOdd = 2.25 \times 10^5$. The spectral lineshape of the reflected signal as a function of the drive and probe detuning can be written as

\begin{equation}
S_{11}=1- \frac{\kappaeEven}{\frac{\kappaEven}{2}+i \deltarpEven+\frac{2\GOM^2}{\gammai+2i(\deltarpEven-(\omegam-\DeltardEven))}},
\label{eq:EIT}
\end{equation}

\noindent where $\deltarpEven \equiv \omegap-\omegacEven$ and $\DeltardEven \equiv \omegacEven-\omegad $. Here, $\kappaeEven$, $\kappaEven$, and $\omegacEven$ denote the parameters of the upper frequency microwave resonance. We extract these parameters first by fitting a Lorentizian lineshape to the microwave resonance when the mechanical response is detuned from the center of cavity ($\DeltardEven \ne \omegam$). Figure \ref{fig:EITfit}b shows the extracted values of $\kappaEven$ as a function of $\omegad$. The external cavity decay rate $\kappaeEven$ is fitted to the constant value of $85.3$~kHz for all the measurements.

Using the extracted values for the microwave resonance we fit the reflection spectrum at the two-photon resonance (`transparency' window) when the pump-probe difference frequency matches that of the capacitively-coupled acoustic mode frequency, $\omegap-\omegad = \omegam$ (See Fig.~\ref{fig:EITfit}c). A particular challenge in modeling the devices of this work is the presence of frequency jitter in the acoustic frequency.  Acoustic frequency jitter tends to blur out the time-averaged measurement of the two-photon resonance peak, and in doing so can lead to a significant underestimation of the coherent electro-mechanical coupling when directly applying Eq.~(\ref{eq:EIT}).  In order to avoid this fitting problem, we first extract the statistical variations in $\omegam$  (frequency jitter) from the linewidth of the mechanical response at the center of transparency window (Fig.\,\ref{fig:EITfit}d). This linewidth contains contributions from electromechanical back-action ($\gammaem = 4\GOM^2/\kappaEven$), intrinsic energy damping of the acoustic mode ($\gammai$), and any pure dephasing (frequency jitter) of the acoustic mode. Using the values ($\GOM/2\pi = 8.9$~kHz and $\gammai/2\pi=68$~Hz) found directly from the time-domain measurements of mechanical ringdown under the same microwave drive conditions, we back out a FWHM frequency jitter value of $\delta\omegam/2\pi = 1.98$~kHz in the spectrum of Fig.\,\ref{fig:EITfit}d.

We next refine our estimates of $\GOM$ and $\gammai$ by fitting a theory curve to the \emph{inverse} of the measured reflection spectrum. Figure \,\ref{fig:EITfit}e shows the spectral shape of the inverse response. The frequency separation between the two peaks in the inverse spectrum can be shown to be very close to the value of $2\GOM$ at two-photon resonance (in the strong coupling limit these two peaks correspond to the hybridized eigenmodes of the coupled system of acoustic and electromagnetic modes).  Importantly, the position of these two peaks in the inverse spectrum are relatively insensitive to acoustic frequency jitter. From this measured frequency separation we find a value of $\GOM/2\pi = 8.4$~kHz, which is in good agreement with the value found from the ring down measurement ($8.9$~kHz).  In the final step of our modeling, we find the value of $\gammai$ by fitting the inverse response using our best estimates of the microwave resonance, parametrically enhanced coupling, and frequency jitter. The value of the best estimate for intrinsic mechanical damping is found to be $\gammai/2\pi=91$~Hz, with a standard deviation of $66$~Hz. In this case the uncertaintity in $\gammai$ is primarily set by the uncertainty in estimating $\kappaEven$, which has been extracted from the curve in Fig.~\ref{fig:EITfit}b.  The slightly elevated value of $\gammai$ from two-tone spectroscopy ($91$~Hz) with respect to the time-domain `in-the-dark' ring-down could be due to the degradation of the mechanical quality factor in presence of pump-indued heating. Nevertheless, we emphasize that the difference between the two values is not statistically significant. We should also note that estimating $\gammai$ in the large cooperativity limit from EIT spectra is inherently challaneging (even in the absence of frequency jitter) due to the saturation in the transparency window peak height.

Using the best estimates of the parameters $\gammai$, $\GOM$, and $\kappaEven$ found from the procedure detailed above, we find the mean value of cooperativity $C = 4\GOM^2/\kappaEven \gammai = 28.5$, with a standard deviation of $\sigma_\mathrm{C} = 7.3$.  This value is consistent with the cooperativity expected from time-domain ringdown measurements.

\section{Frequency jitter analysis}
\label{app:G}

Our analysis of the frequency noise of the breathing mode follows that from Refs.~\cite{Zhang2014,Zhang2015}.  A main takeaway from the referenced work is that frequency fluctuations slower than the decay rate of the the mechanical resonance ($\delta\omega \ll \gammam$), can be considered as slow fluctuations in the mechanical susceptibility leading to slow variation in amplitude and phase of the driven response. Thus an applied near resonanr tone is spectrally broadened around its frequency $\omega_F$. For frequency fluctuations faster than the intrinsic energy decay rate of the mechanical oscillator ($\delta_\omega \gg \gammam$), the driving term quickly loses the memory of the driving frequency, and the mechanical response is broadened over the entire range of the frequency fluctuations.  This results in a similar response spectrum to that of the thermal spectrum, with the only difference being that the amplitude of the signal is dependent on the driving amplitude rather than the temperature of the thermal bath coupled to the mechanical mode. 

Figure~6a of the main text shows the narrowband (slow) and broadband (fast) spectra of the mechanical frequency response in the presence of a weak coherent tone applied near resonance. We denote the area of the narrowband and broadband peaks by $S_\text{nb}$ and $S_\text{bb}$, respectively. We can estimate the broadband component of the frequency jitter of the mechanical oscillator by measuring the area underneath the narrowband and broadband peaks and using the relation
\begin{equation}
\frac{\tilde{\gamma}_m}{\gammam} \approx 1+ \frac{S_{bb}}{S_{\delta}} \left(1-\frac{S_{nb}}{S_{\delta}}\right),
\label{Eq:FreqFluctuation}
\end{equation}
where $\tilde{\gamma}_m$ is the broadened mechanical linewidth and $S_{\delta}$ corresponds to the area underneath the applied coherent tone in the absence of mechanical interaction, measured by detuning the tone frequency away from the mechanical resonance. We have extracted the total linewidth of 6.7 kHz by fitting a Lorentzian to the thermometry response of the mechanical resonator in the absence of any weak drive tones. Using the relation in Eq.~(\ref{Eq:FreqFluctuation}), the broadband (fast) frequency fluctuations contribute to only 15$\%$ of the measured total mechanical linewidth. Back-action cooling $\gammaem$ and intrinsic decay rate $\gammai$ of the mechanical oscillator add up to 29$\%$ of the total linewidth, independently measured using mechanical ring down measurement. We attribute the remaining 58$\%$ of the linewidth to low frequency jitter of the mechanical resonance frequency.

\section{Heating model}
\label{app:H}

The heating curve observed in Fig.6c of the main text fits well to the phenomenological heating model introduced in Ref.~\cite{Meenehan2015}. This model assumes the coupling between the mechanical oscillator and three different thermal baths.  First, the red side-band drive turns on the radiation pressure coupling between the mechanical oscillator and the effective zero-temperature coherent drive with rate $\gammaem$ cooling the mechanical oscillator. Second, the mechanical oscillator itself is coupled to the ambient fridge bath with occupancy $\nmbath$ at an intrinsic rate $\gammai$ and third, it is coupled to a pump induced hot bath with occupancy $\nmpump$ and coupling rate $\gammap$. Moreover, to best capture the dynamics of the heating curve, it is assumed that the hot bath has a finite equilibration time. Thus the red curve fitted to the data, assumes that a fraction of the hot thermal bath turns on almost instantaneously, while the remainder has a slow exponential increase to its steady state value. Thus we can write a simple phenomenological rate equation: 
\begin{equation}
\dot{n}_{m} = -\gamma \nm + \gammap \nmpump(1-\delta_b e^{-\gamma_s t}) + \gammai \nmbath
\end{equation}
where $\gamma = \gammai+\gamma_{em}+\gamma_p$, $\delta_b$ is the slow growing fraction of $\nmpump$ and $\gamma_s$ is the turn-on rate. Assuming a constant $\gammap$, this rate equation has a simple solution of  the form
\begin{equation}
\nm(t)=\nmbath e^{-\gamma t} + n_H (1-e^{-\gamma t})+n_{\delta}(e^{-\gamma_s t}-e^{-\gamma t}) 
\end{equation}
where $n_{\delta}$ and $n_H$ are defined by
\begin{align}
n_{\delta}&=\frac{\gamma_p n_p \delta_b}{\gamma_s-\gamma}\\ n_H &= \gamma^{-1}(\gamma_p n_p + \gammai n_{b,m})
\end{align}



\end{document}